\documentclass{aa}  
\usepackage{graphicx}
\usepackage{url}
\usepackage[normalem]{ulem}
\usepackage{mathrsfs,amssymb,amsmath}
\usepackage[usenames]{color}
\usepackage[varg]{txfonts}
\usepackage{booktabs}
\usepackage{multirow}

\usepackage{bm}

\usepackage[breaklinks, colorlinks, citecolor=blue]{hyperref}
\newcommand{\be}{\begin{equation}}
\newcommand{\ee}{\end{equation}}

\newcommand{\mate}{\mathcal{E}}
\newcommand{\ud}{{\rm d}}

\begin{document} 

\title{Cosmic rays in molecular clouds\\probed by H$_{2}$ rovibrational lines}
\subtitle{Perspectives for the James Webb Space Telescope}


 \author{
          Marco Padovani\inst{1}
          \and
          Shmuel Bialy\inst{2}
          \and
          Daniele Galli\inst{1}
          \and
          Alexei~V. Ivlev\inst{3}
          \and
          Tommaso Grassi\inst{3}
          \and
          Liam~H. Scarlett\inst{4}
          \and 
          Una~S. Rehill\inst{4}
		  \and
		  Mark~C. Zammit\inst{5}
		  \and
		  Dmitry~V. Fursa\inst{4}
          \and
          Igor Bray\inst{4}
          }

   \institute{INAF--Osservatorio Astrofisico di Arcetri, 
              Largo E. Fermi 5, 50125 Firenze, Italy\\
              \email{marco.padovani@inaf.it}
              \and
              Department of Astronomy, University of Maryland, 
              College Park, MD 20742, USA
              \and
              Max-Planck-Institut f\"ur Extraterrestrische Physik, 
              Giessenbachstr. 1,
              85748 Garching, Germany
              \and
              Curtin Institute for Computation and Department of Physics and Astronomy, 
              Curtin University, Perth, Western Australia 6102, Australia
              \and
              Theoretical Division, Los Alamos National Laboratory, Los Alamos, New Mexico 87545, USA
             }


 
  \abstract
{Low-energy cosmic rays (< TeV) play a fundamental role in
the chemical and dynamical evolution of molecular clouds, as they
control the ionisation, dissociation, 
and excitation of H$_{2}$. Their characterisation is 
therefore important both for the interpretation of 
observations and for the development of theoretical
models.
However, the methods used so far for estimating the cosmic-ray
ionisation rate in molecular clouds have several limitations due to
uncertainties in the adopted chemical networks.}
{We refine and extend the method proposed by \cite{Bialy2020}
to estimate the cosmic-ray ionisation rate in molecular clouds by observing 
rovibrational transitions of H$_{2}$ at near-infrared 
wavelengths, which are mainly excited by secondary cosmic-ray electrons. 
}
{Combining models of interstellar cosmic-ray propagation and attenuation in 
molecular clouds with the rigorous calculation of the expected 
secondary electron spectrum and updated H$_{2}$ excitation cross
sections by electron collisions,
we derive the intensity of the four 
H$_{2}$
rovibrational transitions observable in dense, cold gas: $(1-0){\rm O}(2)$, $(1-0){\rm Q}(2)$, 
$(1-0){\rm S}(0)$, and $(1-0){\rm O}(4)$.}
{The proposed method allows the estimation of the 
cosmic-ray ionisation rate for a given observed line intensity
and H$_{2}$ column density.
We are also able to deduce the shape of the 
low-energy cosmic-ray proton spectrum impinging upon the molecular cloud.
We also present a look-up plot and a web-based application that can be used to
constrain the low-energy spectral slope of the interstellar cosmic-ray proton spectrum.
We finally comment on the capability of the James Webb Space Telescope to detect
these near-infrared H$_{2}$ lines, making it possible to derive for the first time
spatial variation of the cosmic-ray ionisation rate in dense gas.
Besides the implications for the interpretation of the 
chemical-dynamic evolution of a molecular cloud, 
it will finally be possible to test competing models
of cosmic-ray propagation and attenuation in the interstellar medium, 
as well as compare cosmic-ray spectra in different Galactic regions.
}
  {}


\maketitle
%
\section{Introduction}
\label{sec:intro}

Cosmic rays (CRs) at sub-TeV energies
play an important role in the energetics and the physico-chemical evolution of star-forming regions. 
Their energy density, of the order of 1~eV~cm$^{-3}$, 
is comparable to that of the Galactic magnetic field, of the cosmic microwave background, and of
the visible starlight \citep{Ferriere2001}.
By ionising molecular hydrogen, the main constituent of molecular clouds, CRs trigger a cascade of chemical reactions leading to the formation of increasingly complex molecules, up to prebiotic species. Furthermore, by determining the ionisation fraction, they regulate the degree of coupling between gas and magnetic field and thus affects 
the collapse timescale of a cloud
\citep[see][for a review]{Padovani+2020}.  

CR particles include electrons, protons and heavier nuclei.
The electron component is revealed by Galactic synchrotron emission, that depends on the strength of the interstellar 
magnetic field 
\citep[e.g.][]{GinzburgSyrovatskii1965,Orlando2018,PadovaniGalli2018,Padovani+2021b}. 
Direct constraints on the spectrum%
\footnote{Also referred to as flux, it represents the number of particles
per unit energy, area, time, and solid angle.} 
of CR electrons can be obtained from synchrotron observations only if the magnetic field strength can be independently estimated by other methods, e.g. by modelling the polarised dust thermal emission
\citep{Alves+2018,Beltran+2019,Sanhueza+2021}.
The proton component of CRs above $\simeq1$~GeV can be constrained through observations of local $\gamma$-ray 
emissivity due to pion decay \citep{Casandjian+2015,Strong+2015,Orlando2018}. However, the results depend on the assumed CR propagation and solar modulation models \citep[see also][for a review]{Tibaldo+2021}. 
At lower energies, between about 3 and 300~MeV, the
local interstellar 
CR spectrum is constrained by 
in situ measurements obtained by the two Voyager spacecrafts \citep{Cummings+2016,Stone+2019}.
Still, the magnetic field direction measured by the Voyager probes did not show
the change expected 
if they were beyond the influence of solar modulation
\citep{GloecklerFisk2015}. Consequently, there is a substantial uncertainty about 
the low-energy CR spectrum.
In addition, fluctuations in the CR spectrum across the Galaxy could be present, 
due to the discrete nature of the CR sources \citep{Phan+2021}.

Several observational techniques provide an estimate of the spectrum
of low-energy CRs in interstellar clouds by determining the ionisation rate, $\zeta_{\rm ion}$, 
i.e. the number of ionisations of hydrogen atoms or molecules per unit time. 
In the diffuse regions of molecular clouds, the CR ionisation rate can be inferred from absorption line studies of H$_{3}^{+}$
\citep{Oka2006, IndrioloMcCall2013},
OH$^{+}$, H$_{2}$O$^{+}$ \citep[see e.g.][]{Neufeld+2010},
and ArH$^{+}$ \citep{NeufeldWolfire2017,Bialy+2019}. 
Even though the method based on H$_{3}^{+}$ absorption lines 
is commonly considered as one of the most reliable, thanks to a particularly 
simple chemistry controlling the H$_{3}^{+}$ abundance \citep{Oka2006}, 
there is a number of 
observational and model limitations that restrict the choice of 
possible target clouds and may introduce significant uncertainties 
in estimating the value of $\zeta_{\rm ion}$. These limitations
include the need 
of having an early-type star in the background, in order to evaluate 
H$_{3}^{+}$ and H$_{2}$ column densities along the same line of sight 
\citep{IndrioloMcCall2012}. 
Furthermore, the value of $\zeta_{\rm ion}$ obtained 
from this method is proportional to the gas volume density and 
therefore is affected by uncertainties in estimating the latter in 
the probed cloud regions 
\citep{JenkinsTripp2001,Sonnentrucker+2007,JenkinsTripp2011,Goldsmith2013}.
Finally, possible strong variations in the 
H$_{3}^{+}$ abundance along the line of sight, caused by uncertainties in 
the local ionisation fraction, which depends on details of 
interstellar UV attenuation in the cloud 
\citep[see][]{NeufeldWolfire2017}, 
may also significantly affect the resulting value of $\zeta_{\rm ion}$.

In denser regions other tracers of $\zeta_{\rm ion}$ 
are used, such as HCO$^{+}$, DCO$^{+}$, and CO in low-mass dense cores \citep{Caselli+1998},
HCO$^{+}$, N$_{2}$H$^{+}$, HC$_{3}$N, HC$_{5}$N, and c-C$_{3}$H$_{2}$ in protostellar clusters
\citep{Ceccarelli+2014,Fontani+2017,Favre+2018},
and more recently H$_{2}$D$^{+}$ and other H$_{3}^{+}$ isotopologues in
high-mass star-forming regions \citep{Bovino+2020,Sabatini+2020}. 
The downside is that the chemistry in these high-density regions is much more complex than in diffuse clouds, requiring 
comprehensive and updated reaction networks. 
In this case, the main source of uncertainty comes from
the 
formation and destruction rates of some species,
which are not well established,
as well as from the poorly constrained amount of carbon and oxygen depletion on dust grains.

We note that the picture is further complicated by the effects of magnetic fields.
If field lines are tangled
and/or the magnetic field strength is not constant, 
as expected in turbulent star-forming regions, CRs can be attenuated more effectively, 
further reducing $\zeta_{\rm ion}$ \citep{PadovaniGalli2011,Padovani+2013,Silsbee+2018}.

Recently, \citet{Bialy2020} developed a new method to estimate the CR ionisation rate 
from infrared observations of rovibrational line emissions of H$_{2}$.
This approach reduces the degree of uncertainty on the determination of $\zeta_{\rm ion}$ with 
respect to the methods listed above, 
as neither chemical networks nor abundances of other secondary species are involved.
These H$_{2}$ rovibrational transitions
are collisionally excited by secondary electrons produced during the propagation of primary CRs. 
In dense molecular clouds most of the H$_2$ 
is in the para form \citep{Bovino+2017,Lupi+2021}.
As we show in Sect.~\ref{sec:linexc}, CRs and UV photons determine the
rovibrational excitation from the $(v,J)=(0,0)$ level to the 
$(v,J)=(1,0)$ and $(1,2)$ levels.
The subsequent radiative decay to the $v=0$ level results in the emission of 
infrared photons at wavelengths of 2--$3~\mu$m (see Table~\ref{tab:H2trans}).
These photons
can be detected by devices such as X-shooter, mounted on the Very Large Telescope (VLT),
the Magellan Infrared Spectrograph (MMIRS), mounted on the Multiple Mirror
Telescope (MMT), see \citet{Bialy+2022},
and by
forthcoming facilities such as the Near Infrared Spectrograph (NIRSpec) on board 
the James Webb Space Telescope (JWST). 
We only consider even-$J$ transitions with $\Delta J=0,\pm2$ (see third column
of Table~\ref{tab:H2trans}) since $|\Delta J|>2$ transitions have negligible 
probability \citep{ItikawaMason2005}.
Besides, odd-$J$ transitions are not frequent in dense molecular clouds
\citep{FlowerWatt1984} as they involve ortho-to-para conversion
due to reactive
collisions with protons.
We also checked that the contribution to the excitation of the 
$(v,J)=(1,0)$ and $(1,2)$ levels by higher vibrational levels is negligible.
For example, the contribution from the $v=2$ level to observed line intensities is less than 
about 5\%.

In this article we refine and extend the method developed by \citet{Bialy2020}, 
taking into account recent advances on the 
calculation of the secondary electron spectrum \citep{Ivlev+2021} 
and updated, accurate H$_{2}$ rovibrational cross sections calculated using
the molecular convergent close-coupling (MCCC) method. 
Thanks to these recent results, we can relax approximations previously made, like, e.g., 
a secondary electron spectrum with
an average energy of about 30~eV
\citep{CravensDalgarno1978} and
a constant ratio of CR excitation and ionisation rates
independent of the H$_{2}$ column density
\citep{GredelDalgarno1995,Bialy2020}. 
In addition, we adopt here the local
CR spectrum as the main parameter of our model. Given the strong dependence
on energy of the cross sections of the processes involved, a spectrum-dependent analysis provides a better parametrisation of the results than a spectrum-integrated quantity like $\zeta_{\rm ion}$,
as assumed by \citet{Bialy2020}.
Assuming a free-streaming regime of CR propagation,
we show that, provided the H$_{2}$ column density is known, the intensity of these 
infrared H$_{2}$ lines 
can constrain both the CR ionisation rate 
and the spectral energy slope of the interstellar
CR proton spectrum at low energies,
This considerably reduces the degree of uncertainty
compared to other methods.

This paper is organised as follows. 
In Sect.~\ref{sec:secondaries} we review the state-of-the-art calculations of the cross sections, and compute an updated energy loss function for electrons in H$_2$, which we use to derive the secondary electron spectrum. 
In Sect.~\ref{sec:zexczion} we calculate the CR excitation rates of H$_2$ and compare them with the CR ionisation rates.
In Sect.~\ref{sec:CRionlookup} we apply the above results to compute the expected observed 
brightness of the H$_{2}$ rovibrational transitions, providing a look-up plot that can be used for a direct estimate of 
the CR ionisation rate and of the low-energy spectral slope of CR protons. 
We also describe the capabilities of JWST 
in detecting the infrared emission of these H$_{2}$
lines.
In Sect.~\ref{sec:conclusions} we summarise our main findings.

\section{Derivation of the secondary electron spectrum}
\label{sec:secondaries}

The brightest H$_{2}$ rovibrational transitions at near-infrared wavelengths,
between 2.22 and 3~$\mu$m, are listed in Table~\ref{tab:H2trans}. 
\begin{table}[!h]
\caption{H$_{2}$ rovibrational transitions.}
\begin{center}
\resizebox{\linewidth}{!}{
\begin{tabular}{cccc}
\toprule\toprule
Transition & Upper level $(v,J)$ & Lower level $(v',J')$ & $\lambda$ [$\mu$m]\\
\midrule
$(1-0){\rm O}(2)$ & (1,0) & (0,2) & 2.63\\
$(1-0){\rm Q}(2)$ & (1,2) & (0,2) & 2.41\\
$(1-0){\rm S}(0)$ & (1,2) & (0,0) & 2.22\\
$(1-0){\rm O}(4)$ & (1,2) & (0,4) & 3.00\\
\bottomrule
\end{tabular}
}
\end{center}
\label{tab:H2trans}
\end{table}%
Their
upper levels, $(v,J)=(1,0)$ and $(1,2)$, 
can be populated very effectively by CR excitation and, to
a lesser extent, by UV or H$_{2}$ formation pumping, respectively
(see~\citealt{Bialy2020} and Sect.~\ref{sec:CRionlookup}). 
CR excitation is dominated by low-energy 
secondary electrons produced during the 
propagation of interstellar CRs, while primary CRs 
(both protons and electrons) provide a negligible contribution to the excitation rate
(see Sect.~\ref{sec:zexczion}).
The rovibrational cross sections of the transitions of interest,
$(v,J)=(0,0)\rightarrow(1,0)$ and $(0,0)\rightarrow(1,2)$
(see second column of Table~\ref{tab:H2trans})
have a maximum around 3--4~eV with a threshold 
at $\sim 0.5$~eV. Therefore, in order to calculate the excitation 
rates, the secondary electron spectrum down to $\sim 0.5$~eV 
needs to be accurately determined.

\citet{Ivlev+2021} developed a rigorous theory for calculating the secondary electron spectrum as a function of 
the primary CR proton spectrum and column density, and 
applied this method to determine the secondary spectrum above the H$_{2}$ 
ionisation threshold ($I=15.44$~eV). In this paper, we extend the calculations of \citet{Ivlev+2021} to lower energies, down to 0.5~eV,
and also include secondary electrons produced by primary CR electrons.
To this goal, the balance equation accounting for all population and depopulation processes of a given energy bin of secondary electrons  
must also include processes occurring at energies $E<I$, such as momentum transfer, rotational excitation $J=0\rightarrow2$ 
and vibrational excitations $v=0\rightarrow1$ and $v=0\rightarrow2$
\citep[see Sect.~4.4 in][for details]{Ivlev+2021}. 
In our previous works \citep[e.g.][]{Padovani+2009,Padovani+2018a,Ivlev+2021}, we made use of the cross 
sections 
summarised by \citet{Dalgarno+1999} 
and the analytical fits of \citet{Janev+2003}.
Recently, a number of theoretical and experimental studies on the H$_{2}$ electronic excitation 
have been published, and in Sect.~\ref{sec:xsections} we comment on the differences with previous studies.

\subsection{Cross sections}
\label{sec:xsections}

\begin{figure*}[!h]
\begin{center}
\resizebox{.9\hsize}{!}{\includegraphics[]{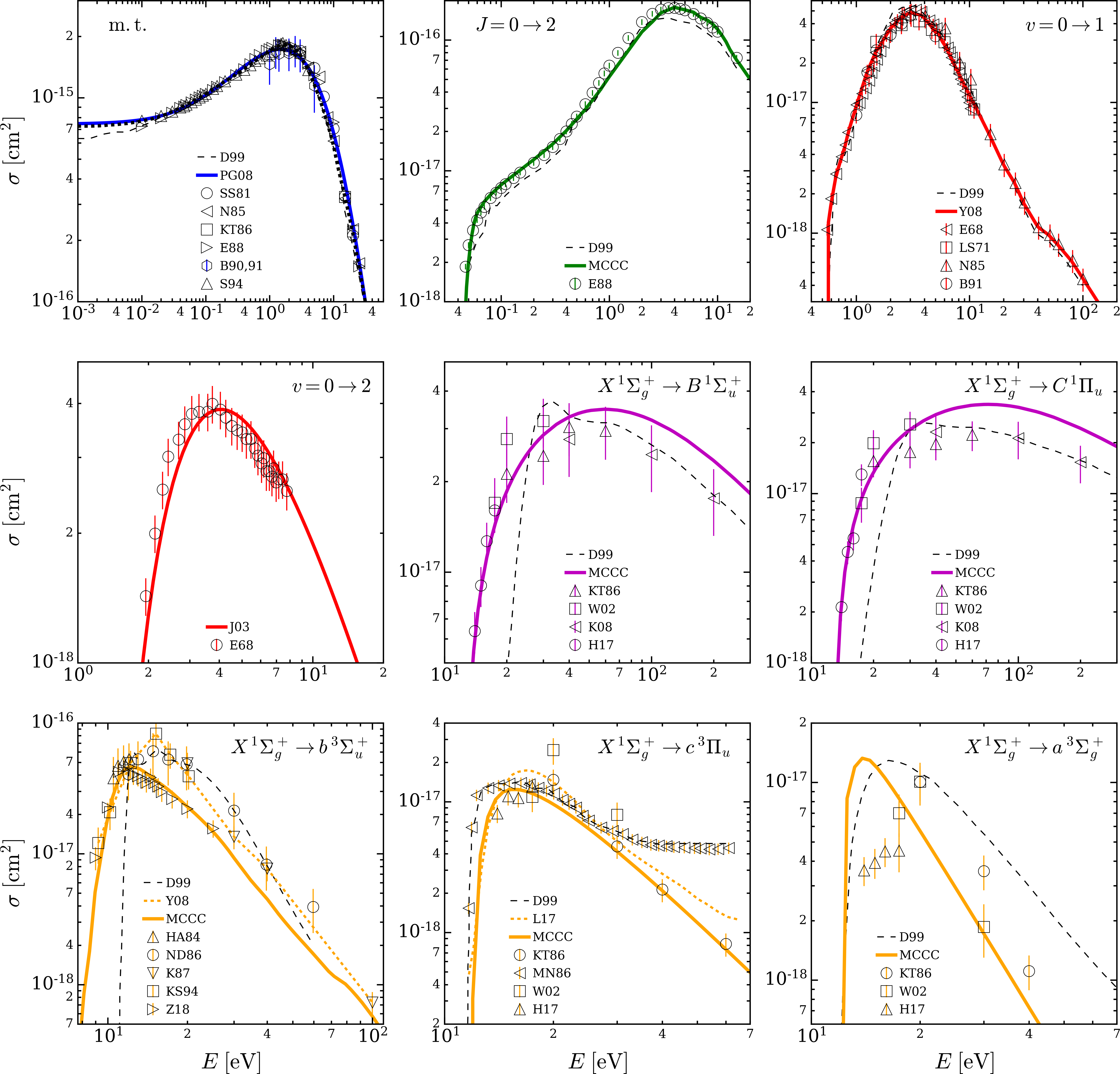}}
\caption{Theoretical and experimental cross sections for electrons colliding with H$_{2}$. The cross sections
used for the calculation of the
energy loss function are displayed as thick lines,
those adopted by \citet{Dalgarno+1999} (D99) by black dashed lines.
From left to right and from top to bottom: 
momentum transfer cross section (``m.t.'') - solid thick blue line \citep[PG08]{PintoGalli2008},
circles \citep[SS81]{ShynSharp1981}, 
left-pointing triangles \citep[N85]{Nishimura+1985}, squares \citep[KT86]{KhakooTrajmar1986},
right-pointing triangles \citep[E88]{England+1988}, hexagons \citep[B90,91]{Brunger+1990,Brunger+1991},
up-pointing triangles \citep[S94]{Schmidt+1994};
rotational transition $J=0\rightarrow2$ - 
solid thick green line (present MCCC calculations), 
circles \citep[E88]{England+1988};
vibrational transition $v=0\rightarrow1$ - 
solid thick red line \citep[Y08]{Yoon+2008},
left-pointing triangles \citep[E68]{Ehrhardt+1968},
squares \citep[LS71]{LinderSchmidt1971},
up-pointing triangles \citep[N85]{Nishimura+1985},
circles \citep[B91]{Brunger+1991};
vibrational transition $v=0\rightarrow2$ - solid thick red line \citep[J03]{Janev+2003},
circles \citep[E68]{Ehrhardt+1968};
$X\,^{1}\Sigma_g^+\rightarrow B\,^1\Sigma_u^+$ and 
$X\,^{1}\Sigma_g^+\rightarrow C\,^1\Pi_u$ singlet transitions - 
solid thick magenta line \citep[MCCC]{Scarlett+2021a},
up-pointing triangles \citep[KT86]{KhakooTrajmar1986},
squares \citep[W02]{Wrkich+2002},
left-pointing triangles \citep[K08]{Kato+2008},
circles \citep[H17]{Hargreaves+2017};
$X\,^{1}\Sigma_g^+\rightarrow b\,^3\Sigma_u^+$ triplet transition - 
dotted orange line \citep[Y08]{Yoon+2008},
solid thick orange line \citep[MCCC]{Scarlett+2021a},
up-pointing triangles \citep[HA84]{HallAndric1984},
circles \citep[ND86]{Nishimura+1986},
down-pointing triangles \citep[K87]{KhakooTrajmar1987},
squares \citep[KS94]{KhakooSegura1994},
right-pointing triangles \citep[Z18]{Zawadzki+2018a};
$X\,^{1}\Sigma_g^+\rightarrow c\,^3\Pi_u$ triplet transition - 
dotted orange line \citep[L17]{Liu+2017},
solid thick orange line \citep[MCCC]{Scarlett+2021a},
circles \citep[KT86]{KhakooTrajmar1986},
left-pointing triangles \citep[MN86]{MasonNewell1986},
squares \citep[W02]{Wrkich+2002},
up-pointing triangles \citep[H17]{Hargreaves+2017};
$X\,^{1}\Sigma_g^+\rightarrow a\,^3\Sigma_g^+$ triplet transition - 
solid thick orange line \citep[MCCC]{Scarlett+2021a},
circles \citep[KT86]{KhakooTrajmar1986},
squares \citep[W02]{Wrkich+2002},
up-pointing triangles \citep[H17]{Hargreaves+2017}.
}
\label{fig:crosssections}
\end{center}
\end{figure*}

In Fig.~\ref{fig:crosssections} we compare available experimental data and earlier theoretical calculations of the main
excitation cross sections  
with the most recent computations adopted in this work (shown by thick solid lines). 
For the electronic excitation cross sections we use the most recent and accurate results
produced using the MCCC method
\citep{Scarlett+2021a}.
These cross sections have already been employing in plasma modelling
\citep{Wuenderlich+2021}, leading to much 
better agreement with measurements compared to the previously-used datasets of \citet{Miles+1972} and \citet{Janev+2003}.
The MCCC results are summarised by 
\citet{Scarlett+2021a} 
and are accessible through a web database.\footnote{\url{https://mccc-db.org/}}

For many transitions, the MCCC method results were found to be in disagreement 
with previously recommended excitation cross sections \citep[e.g.][]{Yoon+2008}.
The most striking difference is for the $X\,^1\Sigma_g^+\rightarrow b\,^{3}\Sigma_{u}^{+}$ transition,
where peak values are twice lower than what recommended
\citep{Scarlett+2017,Zammit+2017}, with 
important consequences on the energy loss function 
(see Sect.~\ref{sec:lossfunction}). On the other hand, recent
experimental results are in perfect agreement with the MCCC calculations
\citep{Zawadzki+2018a,Zawadzki+2018b}.

As for the $X\,^1\Sigma_g^+\rightarrow B\,^{1}\Sigma_{u}^{+}$ and 
$X\,^1\Sigma_g^+\rightarrow C\,^{1}\Pi_{u}$ cross sections, 
there are no recent measurements in the energy region 
close to the cross section peak. We adopt the MCCC calculations because the method is essentially without approximation, aside from the adiabatic-nuclei approximation
which is of no consequence at the energies of interest, where there is disagreement
with older experiments. 
Since for elastic, grand-total, ionisation, and the $X\,^1\Sigma_g^+\rightarrow b\,^{3}\Sigma_{u}^{+}$ 
cross sections the MCCC results are in near-perfect agreement with experiment, 
we adopt the $X\,^1\Sigma_g^+\rightarrow B\,^{1}\Sigma_{u}^{+}$ and 
$X\,^1\Sigma_g^+\rightarrow C\,^{1}\Pi_{u}$ cross sections from 
the MCCC
method as well.
However, close to the energy peak of the singlet cross sections 
the dominant electron loss process is ionisation
(see Fig.~\ref{fig:losselectrons}), 
therefore this difference has no consequences for our purposes.

Recently, \cite{Scarlett+2021b} applied the MCCC method to calculate 
rovibrationally-resolved cross sections for the 
$X\,^1\Sigma_g^+\rightarrow d\,^3\Pi_u$ transition, 
in order to study the polarisation of Fulcher-$\alpha$ fluorescence. 
Here, we apply the same method to calculate cross sections for the rovibrational transitions listed in Table~\ref{tab:H2trans}.

\subsection{Electron energy loss function}
\label{sec:lossfunction}

The quantity that controls the energy degradation of a particle propagating through
a medium is the so-called energy loss function. For electrons
colliding with H$_2$, it is described by%
\footnote{See Eqs. (4) and (5) in \citet{Padovani+2018a} for more details on
the expressions of continuous and catastrophic energy loss processes.}
\begin{eqnarray}\label{eq:lossfunction}
L_{e}(E)&=&\frac{2m_{e}}{m_{\rm H_{2}}}\sigma_{\rm m.t.}(E)E+\sum_{j}\sigma_{{\rm exc},j}(E)E_{{\rm thr},j}+\\\nonumber
&&\int_{0}^{(E-I)/2}\frac{\ud\sigma_{\rm ion}(E,\varepsilon)}{\ud\varepsilon}(I+\varepsilon)\ud\varepsilon+\\\nonumber
&&\int_{0}^{E}\frac{\ud\sigma_{\rm br}(E,E_{\gamma})}{\ud E_{\gamma}}E_{\gamma}\ud E_{\gamma}+KE^{2}\,.
\end{eqnarray}
Terms on the right-hand side represent the contributions of momentum transfer, 
rotational, vibrational, and electronic excitation, ionisation, and bremsstrahlung.
In addition, the last term on the right-hand side
represents synchrotron losses that only depend on the strength of the 
magnetic field in the cloud.
Here, $m_{e}$ and $m_{\rm H_{2}}$ are the electron and H$_2$ mass, respectively,
$\sigma_{\rm m.t.}$ and $\sigma_{{\rm exc},j}$ are the cross section of momentum transfer and excitation of state $j$ summarised in Fig.~\ref{fig:crosssections},
$E_{{\rm thr},j}$ is the corresponding excitation threshold energy, 
$\ud\sigma_{\rm ion}/\ud\varepsilon$ is the differential ionisation cross section \citep{Kim+2000},
where $\varepsilon$ is the secondary electron energy,
and $\ud\sigma_{\rm br}/\ud E_{\gamma}$ is the differential bremsstrahlung cross section
\citep{BlumenthalGould1970}, where $E_{\gamma}$ is the energy of the emitted photon.
Finally, $KE^{2}$ represents synchrotron losses with $K=5\times10^{-38}$~eV~cm$^{2}$ and $E$ in eV
\citep{Schlickeiser2002book}.%
\footnote{Here we assume the relation between the magnetic field strength and the 
volume density given by \citet{Crutcher2012}, $B=B_{0}(n/n_{0})^{\kappa}$, with $B_{0}=10~\mu$G, 
$n_{0}=150$~cm$^{-3}$, and $\kappa=0.5-0.7$. We choose $\kappa=0.5$ to remove the
dependence on $n$ \citep[see][for details]{Padovani+2018a}.}
For typical temperatures ($T\simeq10$~K) and ionisation fractions ($x_{e}<10^{-7}$), Coulomb losses 
are negligible in the energy range of interest \citep{Swartz+1971}.
For clarity, we show the loss functions for the electronic
excitation summed over all the triplet states
($b\,^3\Sigma_u^+$, $a\,^3\Sigma_g^+$, $c\,^3\Pi_u$, 
$e\,^{3}\Sigma_{u}^{+}$, $h\,^{3}\Sigma_{g}^{+}$, $d\,^{3}\Pi_{u}$,
$g\,^{3}\Sigma_{g}^{+}$, $i\,^{3}\Pi_{g}$, and $j\,^{3}\Delta_{g}$) 
and the singlet states
($B\,^1\Sigma_u^+$, $C\,^1\Pi_u$, $EF\,^{1}\Sigma_{g}^{+}$,
$B^{\prime}\,^{1}\Sigma_{u}^{+}$, $GK\,^{1}\Sigma_{g}^{+}$,
$I\,^{1}\Pi_{g}$, $J\,^{1}\Delta_{g}$, $D\,^{1}\Pi_{u}$,
and $H\,^{1}\Sigma_{g}^{+}$).

\begin{figure*}[!h]
\begin{center}
\resizebox{.95\hsize}{!}{\includegraphics[]{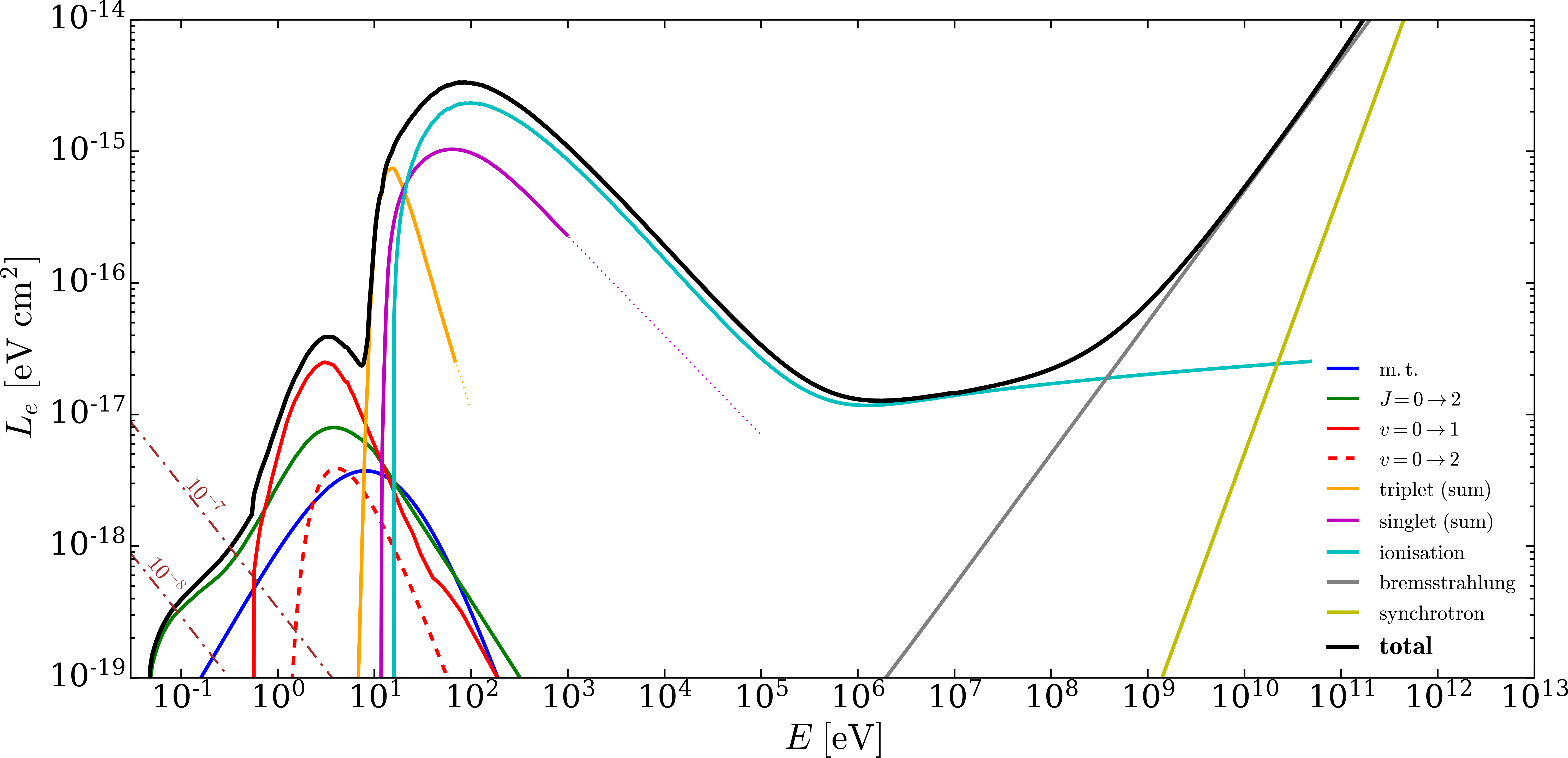}}
\caption{Energy loss function for electrons colliding with H$_2$ 
including the contribution of synchrotron losses
(solid black line). Coloured lines show the 
different components (the following references refer to the papers from which the relative cross sections have been adopted). Momentum transfer (``m.t.'', solid blue; \citealt{PintoGalli2008});
rotational transition $J=0\rightarrow2$ (solid green line; \citealt{England+1988});
vibrational transitions $v=0\rightarrow1$ (solid red line; \citealt{Yoon+2008}) and $v=0\rightarrow2$ (dashed red line; \citealt{Janev+2003});
electronic transitions summed over all the triplet and singlet states
(solid orange and magenta lines, respectively;
\citealt{Scarlett+2021a});
ionisation (solid cyan line; \citealt{Kim+2000});
bremsstrahlung (solid grey line; \citealt{BlumenthalGould1970,Padovani+2018a});
synchrotron (solid yellow line;
\citealt{Schlickeiser2002book,Padovani+2018a}).
Dash-dotted brown lines show the Coulomb losses at 10~K for ionisation fractions, $x_e$, equal to $10^{-7}$ and 
$10^{-8}$ \citep{Swartz+1971}.
}
\label{fig:losselectrons}
\end{center}
\end{figure*}

The resulting energy loss function, $L_e(E)$, shown in Fig.~\ref{fig:losselectrons}, differs in two energy ranges 
from the one adopted in our previous works \citep[e.g.][]{Padovani+2009,Padovani+2018a}, which was based on the cross sections by
\citet{Dalgarno+1999} and data from the 
National Institute of Standards and Technology database\footnote{\url{
physics.nist.gov/PhysRefData/Star/Text/intro.html}}. 
We note that, while \citet{Dalgarno+1999} assume an ortho-to-para ratio of 3:1,
we assume that molecular hydrogen is uniquely in the form of para-H$_{2}$
(see Sect.~\ref{sec:intro}).
The new loss function is a factor of $\simeq 3$ larger between 0.05 and 0.1~eV 
due to the different assumption on temperature and ortho-to-para ratio, and 
is up to 20 times larger in the range $7-12$~eV, 
mainly due to the updated $X\,^1\Sigma_g^+\rightarrow b\,^3\Sigma_u^+$ excitation cross section. For our purposes, the latter difference is especially important for the derivation of the spectrum of secondaries below the H$_{2}$ ionisation threshold.

\subsection{Spectrum of secondary electrons}
\label{sec:secondaryspectra}
We extend the solution of the balance equation, Eq.~(27) in \citet{Ivlev+2021}, down to $0.5$~eV to compute the
secondary electron spectrum at various H$_{2}$ column densities.
We also checked the effect of a change in the composition of the medium, 
including a fraction of He equal to $\simeq 20\%$ \citep[see Table A.1 in][]{Padovani+2018a}. 
However, the additional contribution to the spectrum of secondaries is on average smaller than 3\% 
and we therefore disregard it. For completeness, in Appendix~\ref{app:Heloss}, 
we show the energy loss function for electrons colliding with He atoms
and the cross sections adopted for its derivation.

\begin{figure}[!h]
\begin{center}
\resizebox{\hsize}{!}{\includegraphics[]{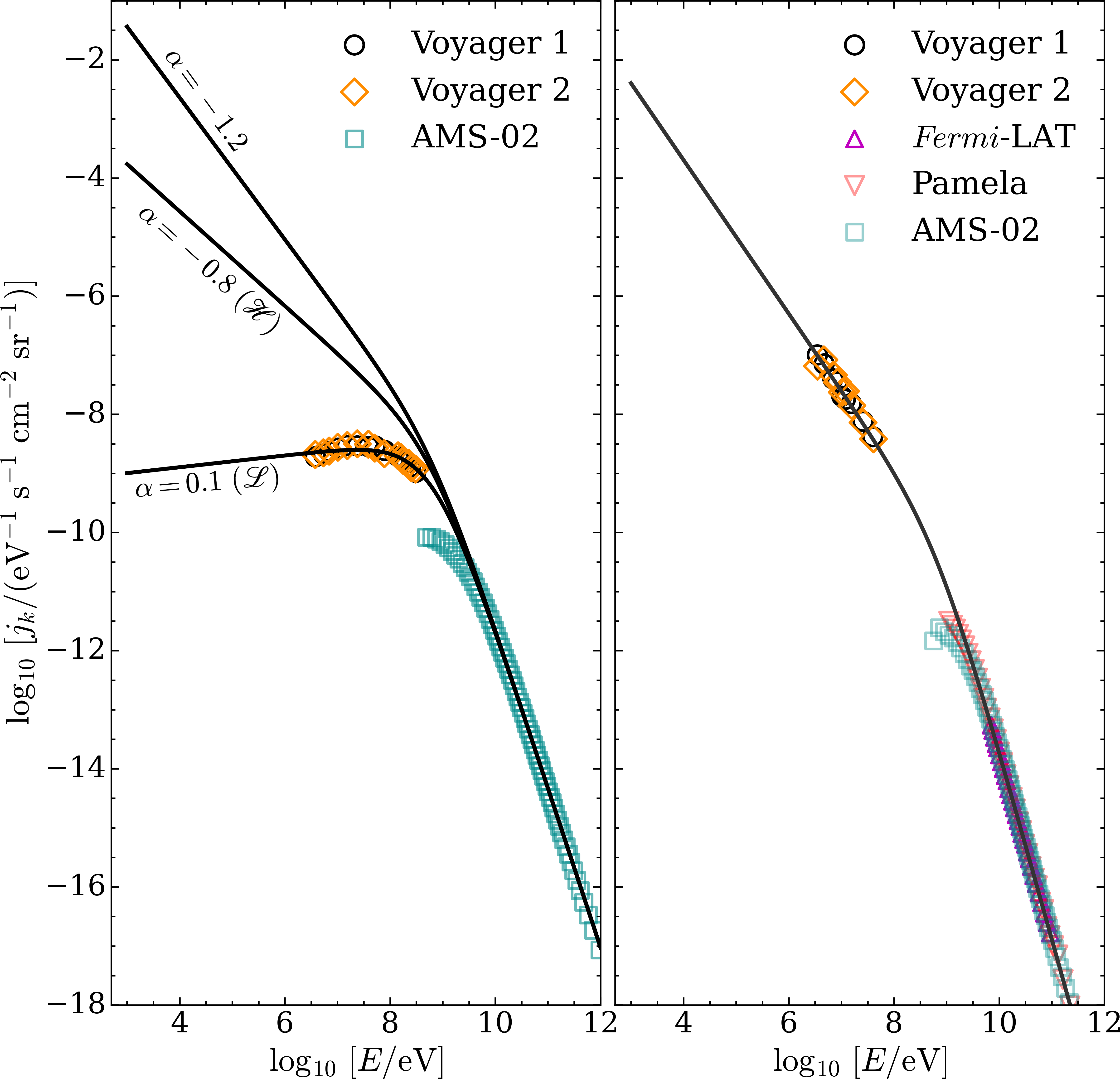}}
\caption{Left panel: CR proton spectrum as a function of the energy for three low-energy
spectral slope: $\alpha=0.1$ (labelled as model $\mathscr{L}$),
$\alpha=-0.8$ (labelled as model $\mathscr{H}$), 
and $\alpha=-1.2$. Right panel: CR electron spectrum as a function of the energy.
Data: Voyager~1 \citep[black circles,][]{Cummings+2016};
Voyager~2 \citep[orange diamonds,][]{Stone+2019};
{\em Fermi}-LAT \citep[magenta up-pointing triangles,][]{Ackermann+2010};
Pamela \citep[pink down-pointing triangles,][]{Adriani+2011};
AMS-02 \citep[cyan squares,][]{Aguilar+2014,Aguilar+2015}.
}
\label{fig:ismspectra}
\end{center}
\end{figure}

For the calculation of the secondary electron spectrum, we assume the analytic form for the  
interstellar CR spectrum from \cite{Padovani+2018a} 
%
\be\label{eq:CRISspectrum}
j_{k}^{\rm IS}(E)=C\frac{E^{\alpha}}{(E+E_{0})^{\beta}}~{\rm eV^{-1}~s^{-1}~cm^{-2}~sr^{-1}}\,,
\ee
where $k=e,p$.
The adopted values of the parameters $C$, $E_{0}$, $\alpha$, and $\beta$ are listed in Table~\ref{tab:CRISparams}. For protons we assume two possible low-energy spectral shapes:
one, with $\alpha=0.1$, reproduces the most recent Voyager~1 and 2 data \citep{Cummings+2016,Stone+2019},
labelled as `low' spectrum $\mathscr{L}$; the other, with $\alpha=-0.8$,
better 
reproduces the average trend of the CR ionisation rate estimated from observations 
in diffuse clouds \citep[][see also Appendix~\ref{app:CRionobstheory}]{Shaw+2008,Neufeld+2010,IndrioloMcCall2012,NeufeldWolfire2017}
and it is labelled as `high' spectrum $\mathscr{H}$.
For the sake of clarity, in this section we consider only these two values of $\alpha$ for protons, 
but in the following sections we allow for 
the whole range of $\alpha$ values, from $-1.2$ to $0.1$
(see left panel of Fig.~\ref{fig:ismspectra}).
As we show in the following sections, most of the parameter space is dominated by the 
ionisation of CR protons and
by the excitation due to secondary electrons. For this reason,
we consider a single parameterisation for 
primary CR electrons
(see right panel of Fig.~\ref{fig:ismspectra}).

\begin{table}[!h]
\caption{Parameters of the interstellar CR electron and proton spectra, Eq.~(\ref{eq:CRISspectrum}), 
where $E$ is in MeV and $C$ is in units of eV$^{-1}$~s$^{-1}$~cm$^{-2}$~sr$^{-1}$.}
\begin{center}
\resizebox{\linewidth}{!}{
\begin{tabular}{lcccc}
\toprule\toprule
Species $k$ & $C$ & $E_{0}~[{\rm MeV}]$ & $\alpha$ & $\beta-\alpha$\\
\midrule
$e$ & $2.1\times10^{18}$ & 710 & $-1.3$ & 3.2\\
$p$ (model $\mathscr{L}$) & $2.4\times10^{15}$ & 650 & $\phantom{-}$0.1 & 2.7\\
$p$ (model $\mathscr{H}$) & $2.4\times10^{15}$ & 650 & $-0.8$ & 2.7\\
\bottomrule
\end{tabular}
}
\end{center}
\label{tab:CRISparams}
\end{table}%

In this work we are interested in H$_{2}$ 
column densities typical of molecular cloud cores ($N_{\rm H_{2}}\lesssim10^{23}$~cm$^{-2}$), 
so we first need to determine how the spectrum of interstellar CRs is attenuated as it propagates within a 
molecular cloud. In this column density regime, it holds the so-called continuous slowing down approximation, 
according to which a CR propagates along a magnetic field line and, each time it collides with an H$_{2}$ molecule, loses a negligible amount of energy compared to its initial energy.
Thus, we assume a free-streaming regime of propagation of CRs
\citep{Padovani+2009}, neglecting their possible 
resonance scattering off small-scale 
turbulent fluctuations, which then may lead to diffusive propagation. 
Therefore, the spectrum of CR particles of species $k$ 
propagated at a column density 
$N_{\rm H_{2}}$, $j_{k}(E,N_{\rm H_{2}})$, can be expressed as a function of the
interstellar CR spectrum at the nominal column density $N_{\rm H_{2}}=0$,
$j_{k}(E_{0},0)$, as
\be
j_{k}(E,N_{\rm H_{2}})=j_{k}(E_{0},0)\frac{L_{k}(E_{0})}{L_{k}(E)}\,,
\ee
where $E$ is the energy of a CR particle with initial energy $E_{0}$ after passing
through a column density $N_{\rm H_{2}}$ given by
\be
N_{\rm H_{2}}=-\int_{E_{0}}^{E} \frac{\ud E}{L_{k}(E)}\,.
\ee
The most updated energy loss function for protons colliding with H$_{2}$ is 
presented in \citet{Padovani+2018a}.

\begin{figure}[!h]
\begin{center}
\resizebox{\hsize}{!}{\includegraphics[]{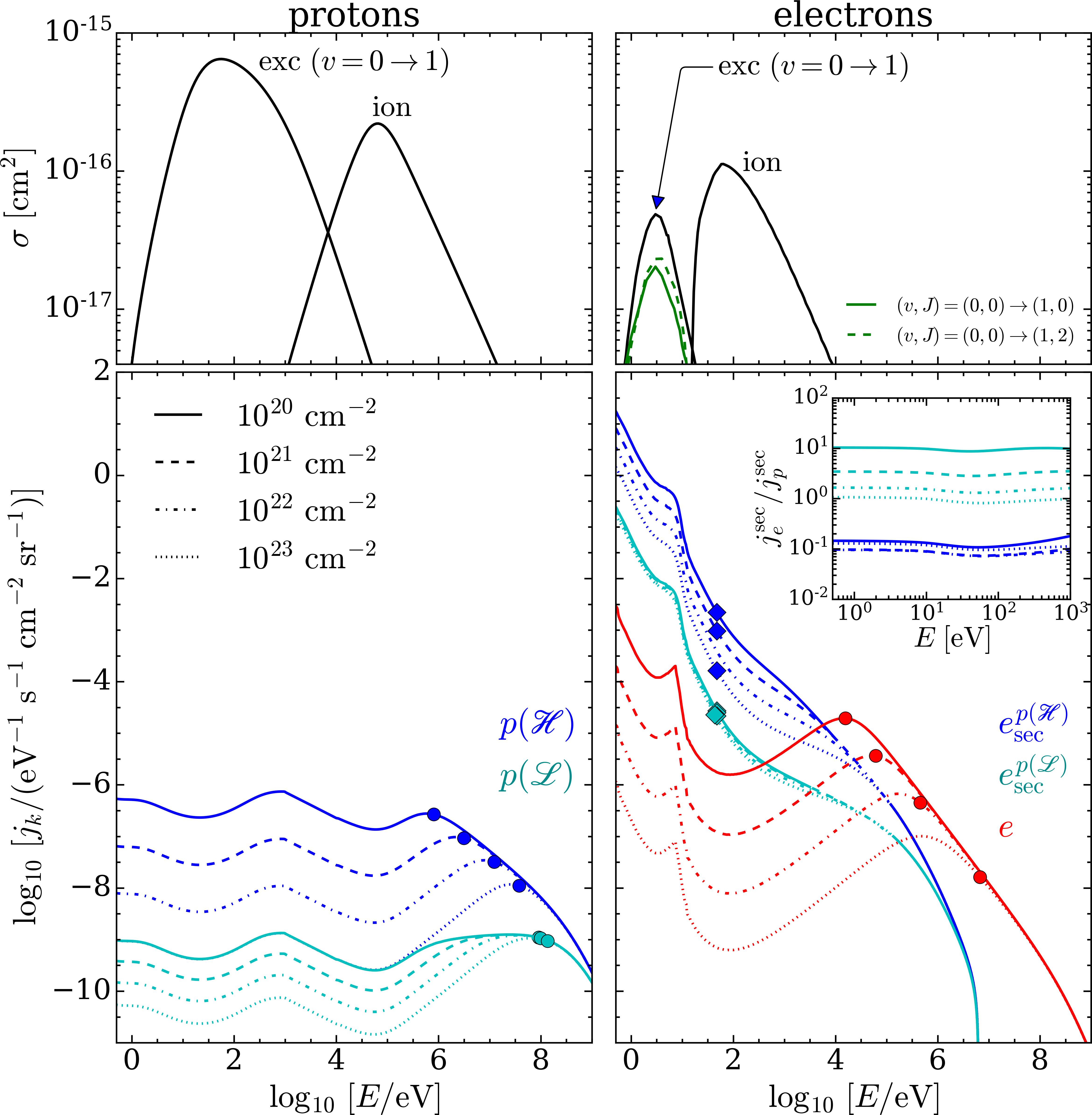}}
\caption{
Upper panels: vibrational excitation, $v=0\rightarrow1$, and ionisation (``ion'') 
cross sections for protons (left plot;
\citealt{TabataShirai2000} and \citealt{Rudd+1992}, respectively) 
and for electrons (right plot;
\citealt{Yoon+2008} and \citealt{Kim+2000}, respectively) colliding with H$_{2}$. 
Solid and dashed green lines show the 
rovibrational cross sections $(v,J)=(0,0)\rightarrow(1,0)$ and $(v,J)=(0,0)\rightarrow(1,2)$, respectively,
from the MCCC calculations.
Lower panels: CR spectra at the column densities  $N_{\rm H_{2}}=10^{20}$, $10^{21}$, $10^{22}$, and $10^{23}$~cm$^{-2}$ as a function of the energy; 
left plot: CR protons (model $\mathscr{L}$ and $\mathscr{H}$; cyan and blue lines, respectively); 
right plot: CR primary electrons ($e$; red lines) and
secondary electrons from CR protons (model $\mathscr{L}$ and $\mathscr{H}$,
labelled as $e_{\rm sec}^{p(\mathscr{L})}$
and $e_{\rm sec}^{p(\mathscr{H})}$, respectively; cyan and blue lines).
The inset shows the ratio between the secondary electron spectra generated by primary CR electrons, $j_{e}^{\rm sec}$, and
by CR protons, $j_{p}^{\rm sec}$ 
(same colour- and line-coding of the main plot).
Solid circles (diamonds) in lower panels 
denote the energies of primary CRs (secondary electrons)
that contribute most to the CR ionisation rate
(see also Appendix~\ref{app:EdzetadE}).
}
\label{fig:spectra}
\end{center}
\end{figure}

The lower left panel of Fig.~\ref{fig:spectra} shows the spectra of CR protons for both models 
$\mathscr{L}$ and $\mathscr{H}$ at four different column densities (from $10^{20}$ to $10^{23}$~cm$^{-2}$). 
The lower right panel shows the corresponding spectra of secondary electrons computed following the procedure
described in
\citet{Ivlev+2021}. We also plot the spectra of CR primary electrons since their contribution to the 
CR ionisation rate is non-negligible when considering proton spectra with
$\alpha\gtrsim-0.4$.
For example, for model 
$\mathscr{L}$,
at $N_{\rm H_{2}}=10^{20}$~cm$^{-2}$ and $10^{21}$~cm$^{-2}$, the contribution of CR primary electrons to the CR
ionisation rate is a factor of 6 and 2 larger, respectively, than that of CR protons. 
At $10^{22}$~cm$^{-2}$ electron and proton ionisation rates are comparable, while at larger
column densities, protons dominate
(see also the lower panel of Fig.~\ref{fig:zexc_zion_vs_N}). 

Additionally, we use the model of \citet{Ivlev+2021} to compute the secondary electron spectrum
from primary CR electrons. As for the latter, we find their contribution 
to ionisation to be non-negligible 
for $\alpha\gtrsim-0.4$ 
(see Sect.~\ref{sec:zexczion}). 
As shown in the lower right panel inset of Fig.~\ref{fig:spectra}, 
the spectrum of secondary electrons produced 
by primary CR electrons is higher by a factor of $\simeq 10$, 3.4, and 1.6 
(at H$_{2}$ column densities of $10^{20}$, $10^{21}$, and $10^{22}$~cm$^{-2}$, respectively)
than that of the secondaries produced by protons for model $\mathscr{L}$.

In contrast to the findings of \citet{CravensDalgarno1978}, 
according to which the spectrum of secondaries 
has an average energy of about 30~eV,
the theory developed by \citet{Ivlev+2021}
predicts that the spectrum of secondaries is distributed over a wide range of energies
(see Appendix~\ref{app:EdzetadE} for more detailed discussion). 

\section{Cosmic-ray excitation and ionisation rates}
\label{sec:zexczion}

The upper panels of Fig.~\ref{fig:spectra} show the excitation and ionisation cross sections 
that we adopt to calculate the corresponding rates,
%
\be\label{eq:zetaind}
\zeta_{k}(N_{\rm H_{2}})=2\pi\ell\int j_{k}(E,N_{\rm H_{2}})\sigma_{k}(E)\ud E\,.
\ee
Here, $\sigma_{k}$ is the excitation or ionisation cross section, and $k$ is the species considered 
(CR protons,
primary CR electrons, and secondary electrons) colliding with H$_{2}$. Assuming a semi-infinite slab geometry, $\ell=1$ for primary CRs
and $\ell=2$ for secondary electrons, since the latter are produced locally and propagate almost isotropically 
\citep[see][]{Padovani+2018b}.
Then, the total ionisation and excitation rates per H$_{2}$ molecule 
are the sum of the individual
contributions given by Eq.~(\ref{eq:zetaind}).

\begin{figure}[!h]
\begin{center}
\resizebox{1\hsize}{!}{\includegraphics[]{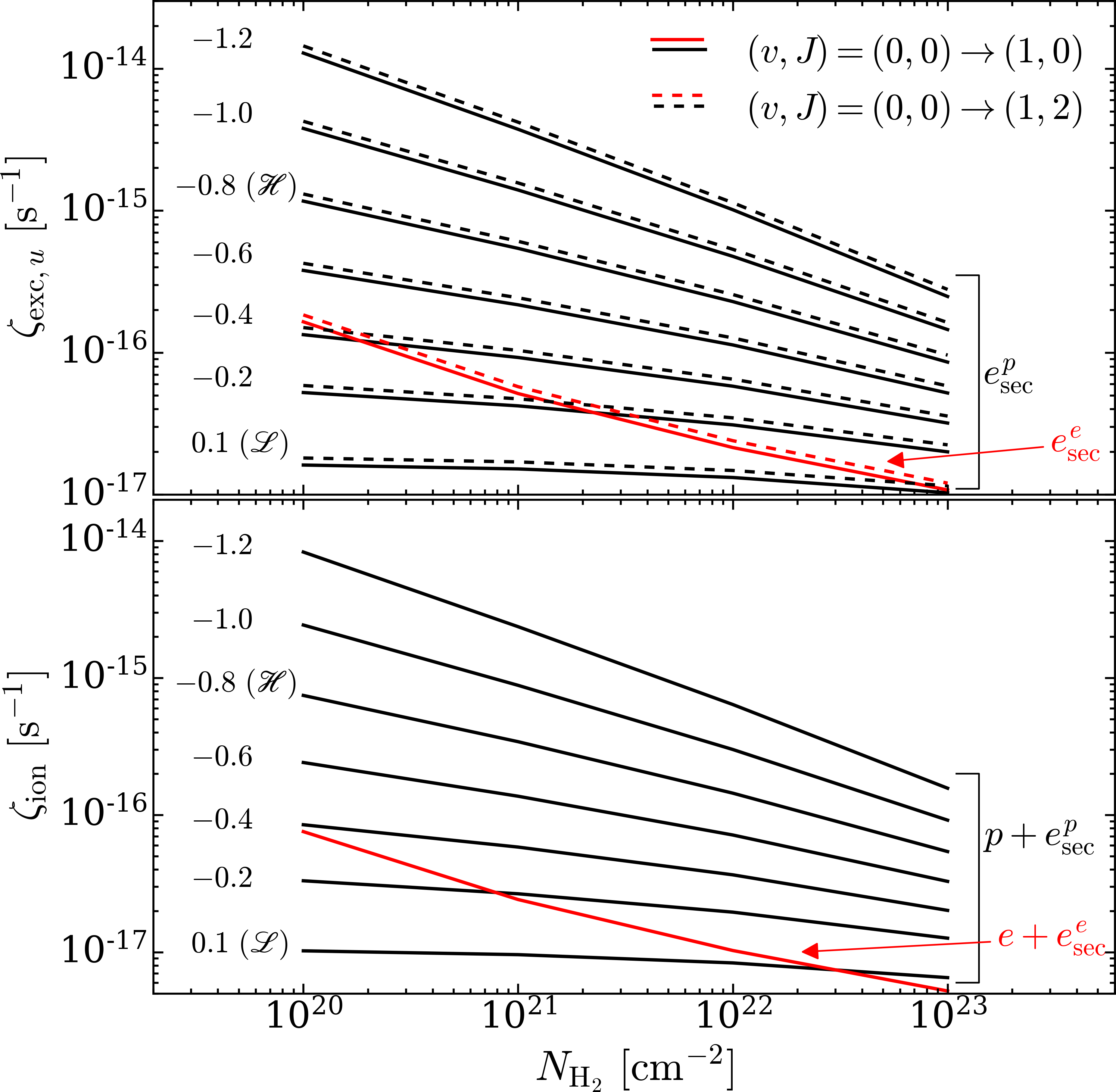}}
\caption{Upper panel: CR excitation rate due to secondary electrons 
as a function of H$_{2}$ column density for the H$_{2}$ rovibrational transitions 
$(v,J)=(0,0)\rightarrow(1,0)$ and $(v,J)=(0,0)\rightarrow(1,2)$ (solid and dashed lines, respectively).
Black (red) lines show the rates due to secondaries produced
by CR protons, $e_{\rm sec}^{p}$ (primary CR electrons, $e_{\rm sec}^{e}$).
Lower panel: CR ionisation rate due to CR protons (solid black lines) and CR electrons 
(solid red line). All the curves include the contribution to ionisation due to the corresponding 
generation of secondary electrons.
Labels on the left in both panels 
denote the low-energy spectral slope (parameter $\alpha$ in Eq.~(\ref{eq:CRISspectrum})).
The cases $\alpha=0.1$ and $\alpha=-0.8$ correspond to model $\mathscr{L}$ and $\mathscr{H}$, respectively. 
}
\label{fig:zexc_zion_vs_N}
\end{center}
\end{figure}

As mentioned at the beginning of Sect.~\ref{sec:secondaries}, we calculate the electron 
excitation rates, $\zeta_{{\rm exc},u}$, 
where $u$ refers to 
the upper $J$ level,
of the rovibrational transitions $(v,J)=(0,0)\rightarrow(1,0)$ 
and $(0,0)\rightarrow(1,2)$.
\citet{Bialy2020} estimated the ratio between excitation and ionisation rates from the excitation probabilities calculated by \citet{GredelDalgarno1995} for 30~eV monoenergetic electrons. Here, we use the H$_{2}$ 
excitation cross sections 
calculated with the MCCC method (see the solid and dashed
green curves in the upper right plot of Fig.~\ref{fig:spectra}), and the spectra of 
primary and secondary electrons computed in the previous section.
The excitation rates for these two transitions
are shown in the upper panel of Fig.~\ref{fig:zexc_zion_vs_N}. 
In particular, we show the excitation rates as a function of the H$_{2}$ column density for different 
low-energy spectral slope, $\alpha$, of the CR proton spectrum. 
We consider not only the models $\mathscr{L}$ and $\mathscr{H}$ described before, with $\alpha=0.1$ and $\alpha=-0.8$, respectively,
but allow $\alpha$ to vary from $-1.2$ to $0.1$. 
As shown by Fig.~\ref{fig:zvsN}, 
$\alpha=-1.2$ gives a CR ionisation rate that represents the upper envelope of the values estimated from observations of diffuse clouds, while $\alpha=-0.8$ 
results in a rate in agreement to average value of the sample.
Values of $\alpha\gtrsim-0.4$ give a rate below the lower envelope of 
observational estimates of $\zeta_{\rm ion}$
in diffuse clouds.%
\footnote{Assuming diffusive propagation of CRs, the case $\alpha=-1.2$ better reproduces 
the average value of $\zeta_{\rm ion}$ in diffuse clouds \citep{SilsbeeIvlev2019}. The 
results of this paper, however, are obtained 
for the free-streaming propagation.}

We also verify that the excitation rate due 
to CR protons is negligible.
Since rotationally-resolved proton-impact cross sections 
are not available, we use the  
vibrational transition $(v,v')=(0,1)$ cross section
summed over all rotational levels recommended by \citet{TabataShirai2000} to obtain 
an upper limit to the H$_{2}$ excitation rate by CR protons. 
Their contribution turns out to be more than three orders of magnitude smaller than that of secondary electrons, therefore it can be safely neglected.
This is because already at column densities of the order of $10^{20}$~cm$^{-2}$, 
protons with energies below about 1~MeV are stopped 
\citep[see Fig.~2 in][]{Padovani+2018a}. 
This implies that the CR proton spectrum is very small at the energies 
where the excitation cross section has its maximum ($\sim100$~eV;
see upper left panel of Fig.~\ref{fig:spectra}).

Excitation by primary CR electrons can also be neglected, since
excitation cross sections peak at $\sim 3$--4~eV, and at these energies the spectra of
secondary electrons generated by protons
are up to $\sim 3$ orders of magnitude higher than the primary CR electron spectrum
(see the middle right panel of Fig.~\ref{fig:spectra}).
However, while primary CR electrons can be neglected, secondary electrons produced by primary CR electrons 
make a non-negligible contribution to the total excitation rate if $\alpha\gtrsim -0.4$ 
(see the red lines in the upper panel of Fig.~\ref{fig:zexc_zion_vs_N}).

The lower panel of Fig.~\ref{fig:zexc_zion_vs_N} shows the ionisation rate due to CR protons and primary CR electrons as a function of column density $N_{\rm H_{2}}$, including the contribution of the corresponding secondary electrons, labelled as $p+e_{\rm sec}^{p}$
and $e+e_{\rm sec}^{e}$, respectively.
Here, the contribution of $e+e_{\rm sec}^{e}$ is not negligible for 
$\alpha\gtrsim-0.4$. In particular, the contribution to ionisation of $e_{\rm sec}^{e}$  
is larger than that of primary CR electrons and increases with H$_{2}$ column density. 
Specifically, the ratio of $\zeta_{\rm ion}$ due to $e_{\rm sec}^{e}$ and to $e$ is
equal to about 1.3, 1.5, 1.7, and 1.9 at $N_{\rm H_{2}}=10^{20}$, $10^{21}$, $10^{22}$, and 
$10^{23}$~cm$^{-2}$, respectively.
Similarly to the excitation rate, primary CR electrons, together with their secondaries, 
determine a lower limit for 
$\zeta_{\rm ion}$ expected from the observations,
independently on the assumed value of $\alpha$. We note, however, that in Fig.~\ref{fig:zvsN} 
there are ionisation rate data below those expected from this  limit. 
This can likely be explained by invoking the presence of highly twisted magnetic field lines, so that
the effective column density passed through by CRs may be much higher than that along the line of sight
\citep{Padovani+2013}. 
Thus the CR spectrum could be strongly attenuated and the corresponding $\zeta_{\rm ion}$ 
may be smaller than predicted. 

\begin{figure}[!h]
\begin{center}
\resizebox{1\hsize}{!}{\includegraphics[]{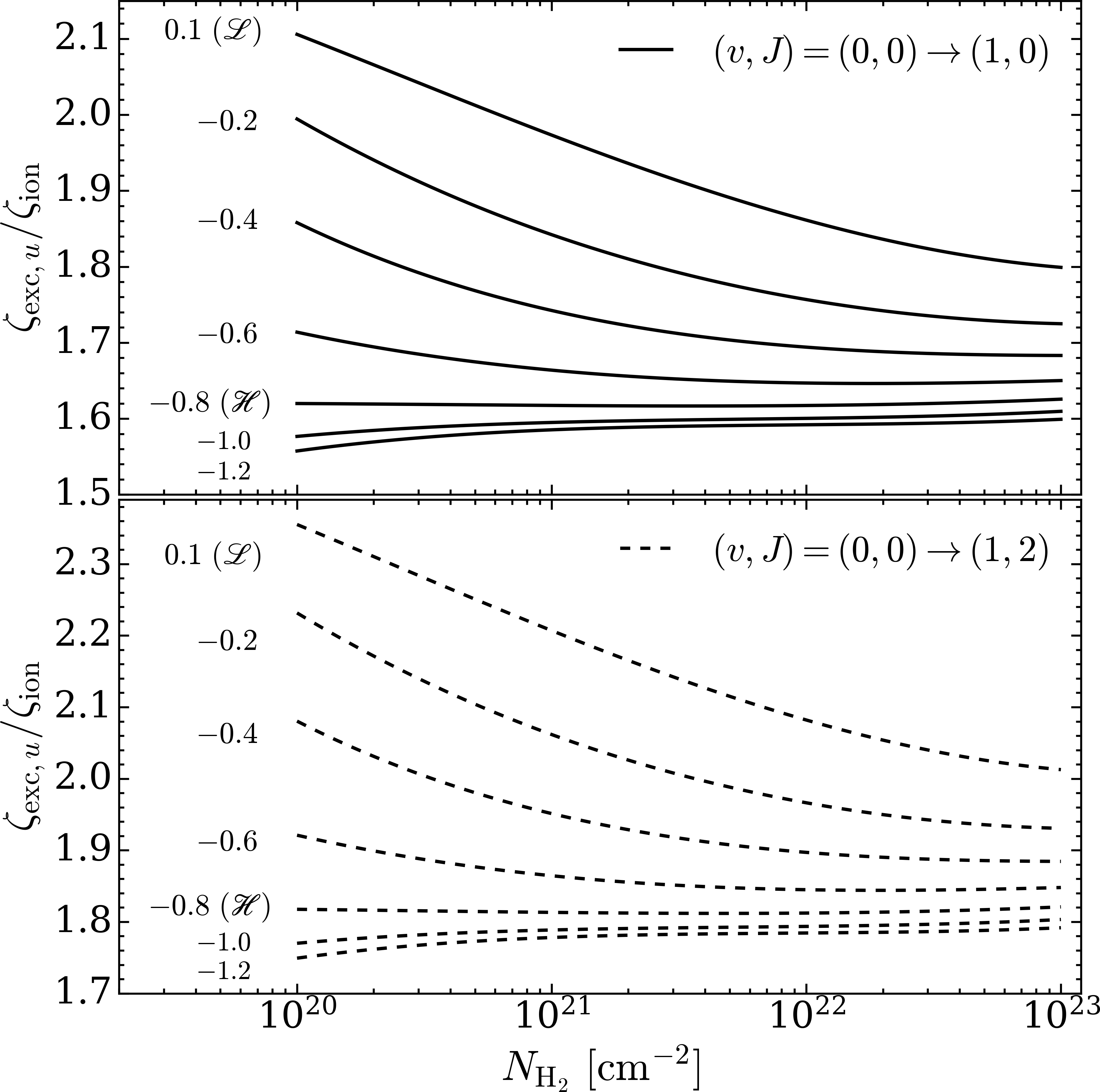}}
\caption{Ratio between the total 
CR excitation and ionisation rates as a function of H$_{2}$ column density
for the H$_{2}$ rovibrational transitions $(v,J)=(0,0)\rightarrow(1,0)$ and $(v,J)=(0,0)\rightarrow(1,2)$
(upper and lower panel, respectively). 
Labels on the left denote the spectral energy slope at low energy (parameter $\alpha$ in Eq.~(\ref{eq:CRISspectrum})).
The cases $\alpha=0.1$ and $\alpha=-0.8$ correspond to model $\mathscr{L}$ and $\mathscr{H}$, respectively. 
}
\label{fig:zexc_zion_ratio}
\end{center}
\end{figure}

Finally, Fig.~\ref{fig:zexc_zion_ratio} shows the ratio between the 
excitation and ionisation rates for the rovibrational transitions 
under consideration. We note that, while in Fig.~\ref{fig:zexc_zion_vs_N} 
the contributions of the various species to excitation and ionisation are shown separately,
here we show the ratio of the total rates.
We find that for increasing H$_{2}$ column densities and 
increasingly negative low-energy spectral slopes $\alpha$,
$\zeta_{{\rm exc},u}/\zeta_{\rm ion}$ tends to an almost constant 
value of $\simeq1.6$ and 1.8, for the 
$(v,J)=(0,0)\rightarrow(1,0)$ and $(v,J)=(0,0)\rightarrow(1,2)$ 
transitions, respectively. For 
$\alpha\gtrsim-0.4$, $\zeta_{{\rm exc},u}/\zeta_{\rm ion}$ reaches larger values 
because of the significant contribution 
of secondary electrons from primary CR electrons to the excitation rate
(see Fig.~\ref{fig:zexc_zion_vs_N}).
\citet{Bialy2020} assumed the ratio between the total excitation rate (summed over
the upper levels) and the ionisation rate to be equal to 5.8.%
\footnote{We remind the reader that \citet{Bialy2020} used the notation
$\zeta_{\rm ex}$ for the total H$_{2}$ excitation to any level.}
Looking at
Fig.~\ref{fig:zexc_zion_ratio}, we see that the $\alpha$- and $N_{\rm H_{2}}$-dependent
value, adding up the excitation rates of the two upper levels considered, ranges from 
3.3 to 4.4. However, results are not directly comparable as in the present work
we also consider the excitation due to secondary electrons from primary CR electrons
and the contribution to ionisation due to both primary CR electrons and their secondaries.

\section{Line excitation}
\label{sec:linexc}

\begin{figure}[!h]
\begin{center}
\resizebox{1\hsize}{!}{\includegraphics[]{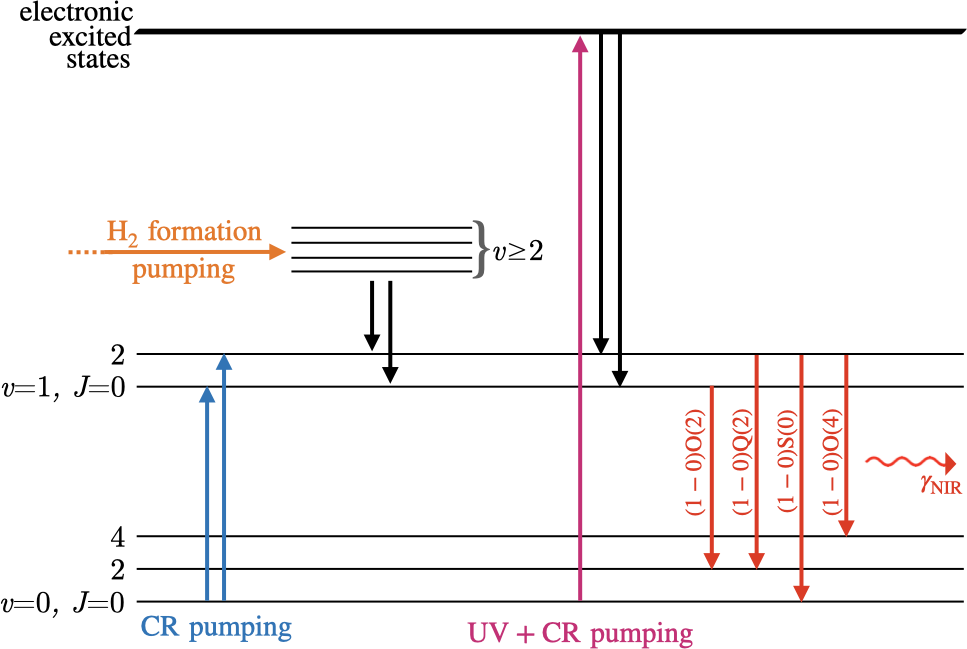}}
\caption{Sketch of the excitation mechanisms contributing to the population of the
$(v,J)=(1,0)$ and $(1,2)$ levels. Direct population is due to CRs (blue arrows) and 
indirect population (black arrows)
occur from the decay of electronic excited states previously populated
by radiative excitation of interstellar UV photons and by collisional excitation by
CRs (magenta arrow) and from the decay of higher vibrational levels ($v\ge2$)
formerly populated as a by-product of the H$_{2}$ formation process (orange arrow).
The four red arrows show the near-infrared (NIR) transitions 
listed in Table~\ref{tab:H2trans}.
We magnify the region of $v=0,1$ levels for clarity.}
\label{fig:sketch-H2-transitions}
\end{center}
\end{figure}

As shown in Fig.~\ref{fig:sketch-H2-transitions}, several mechanisms contribute to the 
population of the $(v,J)=(1,0)$ and $(1,2)$ rovibrational levels.
These levels are populated directly by CRs (blue arrows), more precisely by secondary electrons (see also Sect.~\ref{sec:zexczion}). 
Population also occurs through indirect processes (black arrows). 
Singlet $B\,^{1}\Sigma_{u}^{+}$ and $C\,^{1}\Pi_{u}$ electronic states 
can be excited both radiatively by interstellar UV photons 
and collisionally by CRs (magenta arrow).
The excited electronic states rapidly decay 
into bound 
rovibrational levels of the electronic ground state,
emitting in the Lyman-Werner bands \citep{Sternberg1988}.
A further indirect population process occurs as a side-product of H$_{2}$ formation on 
grains (orange arrow). 
Part of the binding energy is redistributed to the internal excitation of the newly formed H$_{2}$, mainly in the vibrational levels $2\le v\le5$ \citep{Islam+2010}.             
Other fractions of the binding energy 
are converted into dust grain heating and into
kinetic energy of H$_{2}$.
Subsequent decay populates the lower $v=1$ level \citep{BlackVanDishoeck1987}.

We summarise below the equations to compute the expected energy surface
brightness (hereafter ``brightness'') induced by CRs, UV photons, 
and the H$_{2}$ formation process, referring to \citet{Bialy2020}
for further details.
The derivation of the contributions to line intensities by CRs
are similar to those presented in \citet{Bialy2020}. However, we consider the 
more general case where $\zeta_{\rm ion}$ is not constant and thus appears in
the integrals. For more details and limiting cases, see Appendix B 
in \citet{Bialy+2022}.
Equations are given for a generic mixture of hydrogen in atomic and molecular form, 
thus the brightness is a function of the total column density of hydrogen in all its forms, 
$N=N_{\rm H}+2N_{\rm H_{2}}$, where $N_{\rm H}$ and $N_{\rm H_{2}}$ are the atomic and molecular
H$_{2}$ column densities, respectively. 
Since we are mainly interested in molecular cloud cores, 
in the following we assume $N\approx2N_{\rm H_{2}}$. 
Consequently, the fraction of  
molecular hydrogen with respect to the total, 
$x_{\rm H_{2}}=n_{\rm H_{2}}/(n_{\rm H}+2n_{\rm H_{2}})$, 
where $n_{\rm H}$ and $n_{\rm H_{2}}$ are the volume densities of H and H$_{2}$, 
respectively, is set to 1/2.

\subsection{Direct excitation by secondary CR electrons}\label{sec:CRpumping}
The expected brightness of the individual line with upper and lower
levels $u$ and $l$ due to CR excitation is%
\footnote{The brightness has units of energy per unit surface, time,
and solid angle.}
\be\label{eq:Iuldir}
I_{ul}^{\rm dir}(N)= \alpha_{ul}\frac{E_{ul}}{4\pi}\int_{0}^{N}\zeta_{{\rm exc},u}(N')e^{-\tau_{d}(N')}x_{\rm H_{2}}(N')\ud N'\,,
\ee
where $\tau_{d}=\sigma_{d}N$ is the optical depth for dust extinction and 
$\sigma_{d}\approx4.5\times10^{-23}$~cm$^{2}$ is the cross section per hydrogen nucleus averaged over
$2-3~\mu$m \citep{Draine2011book,Bialy2020}.
Here, $\alpha_{ul}$ 
is the probability to decay to state $l$ given state $u$ is excited
and
$E_{ul}$ is the transition energy (see Table 1 in \citealt{Bialy2020}).
We note that H$_{2}$ self-absorption is negligible with respect to
the absorption by dust at these wavelengths.

\subsection{Indirect excitation by interstellar and CR-induced UV photons}
\label{sec:UVpumping}
The expected brightness due to interstellar UV photons and CR-excited Lyman-Werner (LW) transitions is
\be\label{eq:IulLW}
I_{ul}^{\rm LW}(N)=f_{ul}^{\rm LW}\frac{\bar E_{\rm UV}}{4\pi}[\mate^{\rm LW}_{\rm ISRF}(N)+\mate^{\rm LW}_{\rm CR}(N)]
\ee
where 
\be\label{eq:FISRF}
\mate^{\rm LW}_{\rm ISRF}(N)=\int_{0}^{N}P_{0}\chi_{a}(N')x_{\rm H_{2}}(N')\ud N'
\ee
and
\be\label{eq:FCR}
\mate^{\rm LW}_{\rm CR}(N)= \mate^{\rm LW}_{\rm CR,0}(\omega,R_{\rm V})%
\left[\frac{\zeta_{\rm ion}(N)}{10^{-17}~{\rm s}^{-1}}\right]
\ee
are the total UV emission rates per unit area
resulting from the decay of the $B\,^{1}\Sigma_{u}^{+}$ and
$C\,^{1}\Pi_{u}$ states excited by the UV interstellar
radiation field (ISRF) and CRs, respectively.
Here, $P_{0}\simeq9D_{0}$ is the unattenuated UV pumping rate
\citep{Bialy2020},
$D_{0}=2\times10^{-11}G_{0}$~s$^{-1}$
is the unattenuated photodissociation rate
\citep[][assuming a semi-infinite slab geometry]{DraineBertoldi1996}, 
$G_{0}$
is the far-UV radiation field in Habing units 
\citep{Habing1968}, and
$\chi_{a}(N)=f_{\rm sh}\exp[-\tau_{g}(N)]$ 
accounts for the self-shielding effect of H$_{2}$
and dust extinction.
The H$_{2}$ self-shielding function is given by
\citet{DraineBertoldi1996} 
\be
f_{\rm sh}=\frac{a_{1}}{(1+x/b_{5})^{2}}+%
\frac{a_{2}}{\sqrt{1+x}}\exp\left(-a_{3}\sqrt{1+x}~\right)\,,
\ee
where 
$a_{1}=0.965$, $a_{2}=0.035$, $a_{3}=8.5\times10^{-4}$,
$x=N_{\rm H_{2}}/(5\times10^{14}~{\rm cm^{-2}}$),
and $b_{5}$ is the absorption-line Doppler parameter normalised to $10^{5}~{\rm cm~s^{-1}}$.
We set $b_{5}=2$ as in \citet{BialySternberg2016}.
Finally, $\tau_{g}=\sigma_{g}N$,
where $\sigma_{g}=1.9\times10^{-21}$~cm$^{2}$
is the average value of the far-UV dust grain absorption 
cross section for solar metallicity 
\citep{Draine2011book}.
We recall that we assume $N=2N_{\rm H_{2}}.$
The total CR-induced UV emission rate per unit area, $\mate^{\rm LW}_{\rm CR}$, is given by 
\citet{Cecchi-PestelliniAiello1992} \citep[see also][]{Ivlev+2015}, where
\be\label{eq:FCR0}
\mate^{\rm LW}_{\rm CR,0}(\omega,R_{\rm V})\simeq\frac{960}{1-\omega}\left(\frac{R_{\rm V}}{3.2}\right)^{1.5}~{\rm cm^{-2}~s^{-1}}\,.
\ee
Here, $\omega$ is the dust albedo at UV wavelengths and $R_{\rm V}$ is a measure
of the extinction at visible wavelengths \citep{Draine2011book}.
Finally, $\bar E_{\rm UV}\simeq1.82$~eV is the effective transition energy and
$f_{ul}^{\rm LW}$ is the relative emission of the transition from
level $u$ to level $l$ (see \citealt{Sternberg1988} and Table 1 in \citealt{Bialy2020}).
We find that $\mate^{\rm LW}_{\rm CR}\ll \mate^{\rm LW}_{\rm ISRF}$ at any column density, thus 
we can safely neglect the contribution of the 
term in Eq.~(\ref{eq:FCR}) to $I_{ul}^{\rm LW}$ (Eq.~(\ref{eq:IulLW})).

\subsection{Indirect excitation from H$_{2}$ formation}\label{sec:H2pumping}
The expected brightness due to H$_{2}$ formation pumping
is
\be\label{eq:IulH2f}
I_{ul}^{\rm f}(N)= f_{ul}^{\rm f}\frac{\bar E_{\rm f}}{4\pi}%
[\mate^{\rm f}_{\rm ISRF}(N)+\mate^{\rm f}_{\rm CR}(N)]\,,
\ee
%
%
where the two terms on the right-hand side 
represent the total emission rates per unit area due to the destruction of H$_{2}$
by interstellar UV photons and by CRs, respectively. They are given by
\be
\mate^{\rm f}_{\rm ISRF}(N)=\int_{0}^{N}D_{0}\chi_{a}(N')x_{\rm H_{2}}(N')\ud N'
\ee
and
\be
\mate^{\rm f}_{\rm CR}(N)=\int_{0}^{N}(y+\Phi_{\rm diss})\zeta_{\rm ion}(N)e^{-\tau_{d}(N)}x_{\rm H_{2}}(N)\ud N\,.
\ee
%
Here, 
$\bar{E}_{\rm f}\simeq1.3$~eV
corresponds to the excitation of the $v=4$ level
\citep{Islam+2010},
the relative emission
of the transition from level $u$ to level $l$, 
$f_{ul}^{\rm f}$,
is determined by the formation excitation pattern
(see \citealt{BlackVanDishoeck1987} and Table 1 in \citealt{Bialy2020}),
$y\simeq2$
accounts for additional removal of H$_{2}$ by H$_{2}^{+}$ 
in predominantly molecular gas \citep{BialySternberg2015},
and
$\Phi_{\rm diss}\simeq0.7$ accounts for the fact that H$_{2}$ 
can also be destroyed through dissociation in addition to ionisation
\citep{Padovani+2018b}.

\section{A look-up plot for $\zeta_{\rm ion}$ and $\alpha$}
\label{sec:CRionlookup}
\begin{figure*}[!h]
\begin{center}
\resizebox{1\hsize}{!}{\includegraphics[]{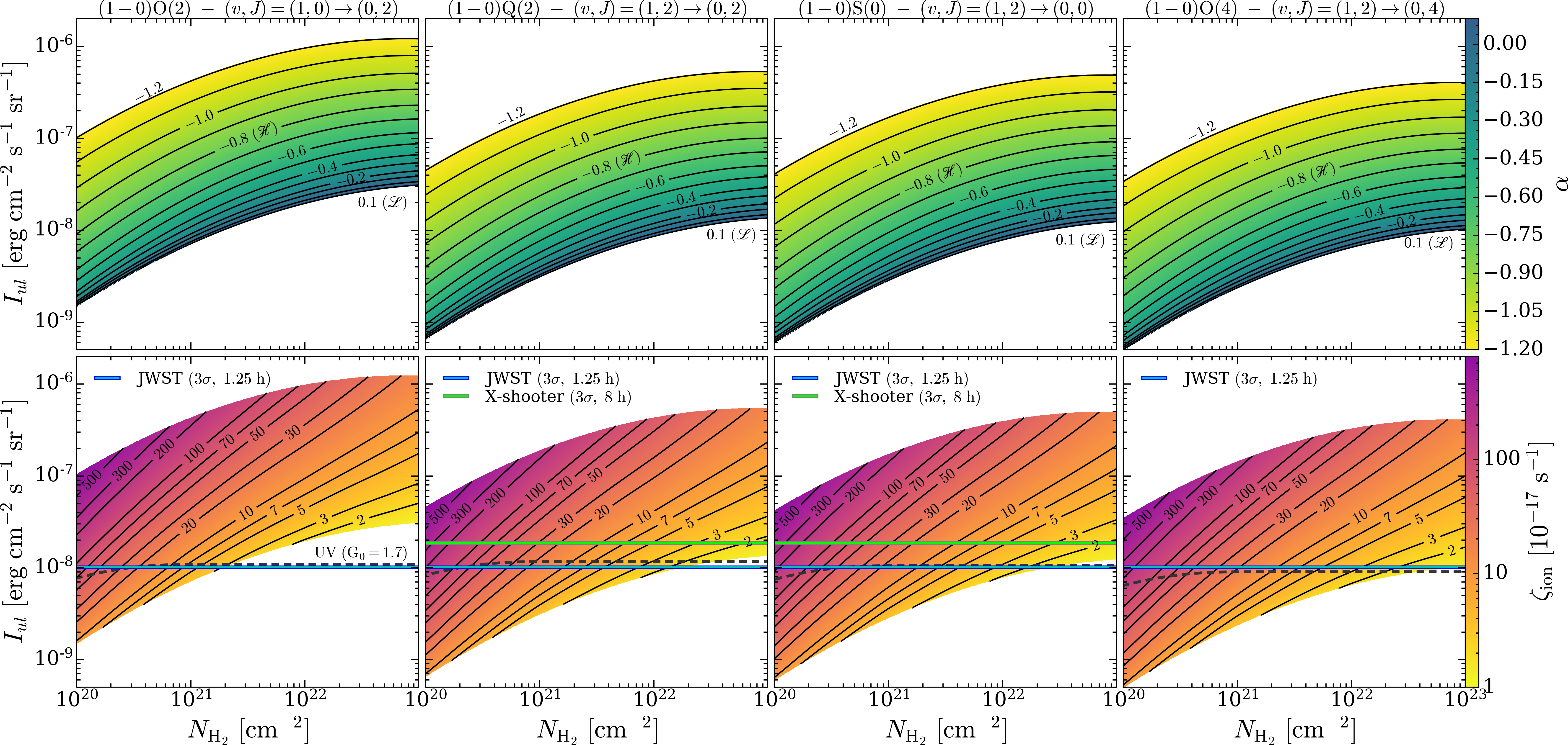}}
\caption{Maps of the 
the low-energy spectral slope ($\alpha$, upper row)
and of the CR ionisation rate ($\zeta_{\rm ion}$, lower row)
as a function
of the energy surface brightness expected by direct CR excitation
for the four H$_{2}$ rovibrational transition listed in Table~\ref{tab:H2trans}
and of the H$_{2}$ column density.
We note that we assume $N=2N_{\rm H_{2}}$.
The dashed black lines show the expected brightness due to indirect excitation by 
interstellar UV photons 
for a far-UV radiation field with $G_{0}=1.7$ \citep{Draine2011book}.
Solid black lines show the iso-contours of 
$\alpha$ (upper panels) and 
$\zeta_{\rm ion}$ in units of $10^{-17}$~s$^{-1}$ (lower panels).
Solid blue horizontal lines show the JWST sensitivity for a signal-to-noise ratio of 
3 over 1.25~h of integration, adding up the signal over 50 shutters.
Solid green horizontal lines show the X-Shooter sensitivity for a signal-to-noise ratio
of 3 over 8~h of integration, adding up the signal over the whole slit.}
\label{fig:I_O2Q2S0O4}
\end{center}
\end{figure*}

Figure~\ref{fig:I_O2Q2S0O4} shows the expected brightness for direct excitation by secondary electrons and
indirect excitation by UV photons,
for the four rovibrational transitions listed in Table~\ref{tab:H2trans}.
The contribution of H$_{2}$ formation pumping is not shown because it is smaller by a
factor 20 to 200 than that of direct CR excitation (depending on the transition considered), 
so it can be safely neglected. 
A similar conclusion was obtained by \citet{Bialy2020}, see their Fig.~1. 
For a UV field equal to the mean interstellar field ($G_{0}=1.7$), 
CRs dominate the excitation if the observed brightness is larger than about
$10^{-8}$~erg~cm$^{-2}$~s$^{-1}$~sr$^{-1}$,
for column densities higher than about a few times $10^{21}$~cm$^{-2}$,
depending on the transition.

Figure~\ref{fig:I_O2Q2S0O4} provides a look-up plot for 
a direct estimate of $\zeta_{\rm ion}$, 
overcoming the uncertainties of other observational methods (see Sect.~\ref{sec:intro}).
We also note that the simultaneous observation of several transitions provides more stringent
constraints on $\zeta_{\rm ion}$.
With this diagram, it is also possible to determine the slope of the CR proton spectrum 
at low energies and to compare it to measurements by the Voyager 
spacecrafts ($\alpha=0.1$).
We remind the reader that, using our model for CR propagation and generation
of secondary electrons, we relate the CR ionisation rate in the cloud to the
unattenuated CR proton spectrum impinging upon the cloud, which is characterised by a 
low-energy spectral slope $\alpha$ (see Sect.~\ref{sec:secondaryspectra}).
In order to facilitate the usage of Fig.~\ref{fig:I_O2Q2S0O4}, we 
have developed a publicly available web-based application%
\footnote{\url{https://cosmicrays-h2rovib.herokuapp.com}} that 
allows a more accurate value of the ionisation rate and of the low-energy spectral
slope
to be obtained, 
given the line brightness and the corresponding column density.

The expected brightness in Fig.~\ref{fig:I_O2Q2S0O4} applies to
typical interstellar UV fields ($G_{0}=1.7$)
and to the average interstellar CR spectrum based on measurements in the solar neighbourhood.
However, different regions of dense gas are likely to be dominated by local conditions,
such as perturbations in the magnetic field structure or shocks.
This could cause variations in the shape of the CR spectrum.
For example, in the vicinity of protostars, 
the UV field can be much more intense ($G_{0}\gg1$), 
especially close to shocks 
\citep[e.g.][]{HollenbachMcKee1989,Karska+2018}. 
However, in the same shocks,
e.g.
along a protostellar jet or on the surface of a protostar, it is also possible to locally 
accelerate CRs \citep{Padovani+2015,Padovani+2016,GachesOffner2018,Padovani+2021a}, 
and therefore even more intense H$_{2}$ 
lines should be observed. 
Consequently, this technique could also be used to further confirm 
the enhanced ionisation triggered by local CRs expected in star-forming regions.

\citet{Bialy2020} showed that X-shooter can be used to observe the $(1-0){\rm Q}(2)$ and 
$(1-0){\rm S}(0)$ lines of H$_2$.
One of the limitations of X-shooter is the small size of the slits ($11''\times0.4''$), 
which allow only a small portion of a starless core to be observed,
whose typical size is of the order of 0.1~pc.
Unfortunately, the brightest H$_{2}$ rovibrational line, $(1-0){\rm O}(2)$, 
cannot be observed from the ground due to atmospheric absorption, 
while the $(1-0){\rm O}(4)$ transition falls outside the range of frequencies observable by X-shooter.
\citet{Bialy+2022} recently employed this 
new method for the determination of $\zeta_{\rm ion}$ using  
MMIRS mounted on MMT,
obtaining for five dense molecular clouds
upper limits on the $(1-0){\rm S}(0)$ transition and the CR ionisation rate
(of the order of $10^{-16}$~s$^{-1}$, 
see also Appendix~\ref{app:CRionobstheory}).
These observations successfully confirmed the validity
of this method,
setting the ground for future observations 
with JWST.

The NIRSpec instrument mounted on JWST turns out to be the crucial facility
for observing these H$_{2}$ infrared lines. 
Indeed, 
in addition to making it possible to observe all four H$_{2}$ transitions in Table~\ref{tab:H2trans},
NIRSpec used in multi-object spectroscopy 
mode 
provides slits with an angular extent of $3.4'$ and a width of $0.27''$.
Adding up the signal over 50 shutters,%
\footnote{Each shutter has a
size of approximately $0.53''\times0.27''$.} the $3\sigma$ threshold 
is achieved in only 1.25~h of observation \citep[see][for more details]{Bialy+2022}.
Given the high spatial resolution, 
this also means that for a starless core such as Barnard 68, at a distance of 125~pc 
\citep{deGeus1989}, it is possible to obtain about 10 independent estimates of the
brightness, and hence of
$\zeta_{\rm ion}$, across the core.

Therefore, in principle it will be possible to obtain for the 
first time the spatially-resolved distribution of the CR ionisation rate 
in a starless core and not a single estimate of $\zeta_{\rm ion}$ as obtained through 
the methods described in Sect.~\ref{sec:intro}.
An important consequence is the possibility of testing the
presence of
a gradient of $\zeta_{\rm ion}$, 
predicted by models of attenuation of the interstellar 
CR spectrum as CRs propagate through a molecular cloud
\citep{Padovani+2009,PadovaniGalli2011,Padovani+2013,Padovani+2018a,Silsbee+2018,SilsbeeIvlev2019}, or whether $\zeta_{\rm ion}$ 
is nearly spatially uniform, in case CRs are accelerated inside 
a cloud 
by magnetic reconnection events \citep{Gaches+2021}. 

Lower panels of Fig.~\ref{fig:I_O2Q2S0O4} 
also show the $3\sigma$ limit for 8~h of integration with X-shooter and
1.25~h of integration with JWST.

\section{Conclusions}
\label{sec:conclusions}

In this paper we presented a 
detailed numerical method to test and extend the analytic model by
\citet{Bialy2020}. Our modelling allows a robust estimate of the CR ionisation rate, 
$\zeta_{\rm ion}$,
and of the low-energy spectral slope of the CR proton spectrum, $\alpha$, in dense molecular clouds
from the observation of photons emitted at near-infrared
wavelengths 
by the decay of rovibrational levels of 
molecular hydrogen.
This technique 
allows to quantify $\zeta_{\rm ion}$ independently on any chemical network.

In a molecular cloud, when sufficiently far away from UV sources such as a protostar, the excitation of the 
$(v,J)=(1,0)$ and $(1,2)$ levels of H$_{2}$ is dominated by secondary CR electrons.
It is traditionally assumed that the spectrum of secondary CR electrons 
has an average energy of about 30~eV 
\citep{CravensDalgarno1978}. However, 
the spectrum of secondary electrons produced during 
the propagation of primary CRs (both protons and electrons)
can be computed accurately at the energies of interest
\citep{Ivlev+2021}. In 
addition, rigorous theoretical calculations of electron-impact
excitation cross sections of 
rovibrational levels of H$_{2}$ are now available
\citep{Scarlett+2021a}.

Finally, following \cite{Bialy2020}, we computed the expected 
brightness for the H$_{2}$ transitions listed in 
Table~\ref{tab:H2trans}.
We then presented a look-up plot, accompanied by an interactive on-line tool, that allows to obtain a straightforward 
estimate of $\zeta_{\rm ion}$ and $\alpha$, 
given the brightness of an H$_{2}$ transition and 
the corresponding column density.
The feasibility of this type of observation was recently verified by \citet{Bialy+2022} using the 
spectrograph 
MMIRS mounted on the MMT, obtaining upper limits for $\zeta_{\rm ion}$
in five dense molecular clouds. 
However, it will be the new generation instrument JWST 
that will allow the application of 
this technique with a great improvement in terms of sensitivity 
and spatial resolution, leading in principle to an actual line detection. 
In fact, while today the current methods provide a single
CR ionisation rate estimate per observed source, 
JWST
will allow to derive the CR ionisation rate profile through a starless core with a single pointing. 
For example, with 1.25~h of observation with JWST, 
up to about 10 independent $\zeta_{\rm ion}$ estimates 
can be derived with a $3\sigma$ sensitivity. 
In addition to having major implications on the interpretation 
of the chemical composition of a molecular cloud and its dynamical 
evolution, the determination of $\alpha$ and of the profile of $\zeta_{\rm ion}$ 
will also make it possible to test the predictions of 
models of CR propagation in molecular clouds 
\citep[e.g.][]{EverettZweibel2011,MorlinoGabici2015,SilsbeeIvlev2019,Padovani+2020,Gaches+2021}.


\begin{acknowledgements}
The authors thank Jonathan Tennyson for insightful comments on cross sections.
\end{acknowledgements}

\bibliographystyle{aa} 
\bibliography{mybibliography-bibdesk.bib} 

\appendix

\section{Energy loss function for electrons in helium}
\label{app:Heloss}
\begin{figure}[!h]
\begin{center}
\resizebox{1\hsize}{!}{\includegraphics[]{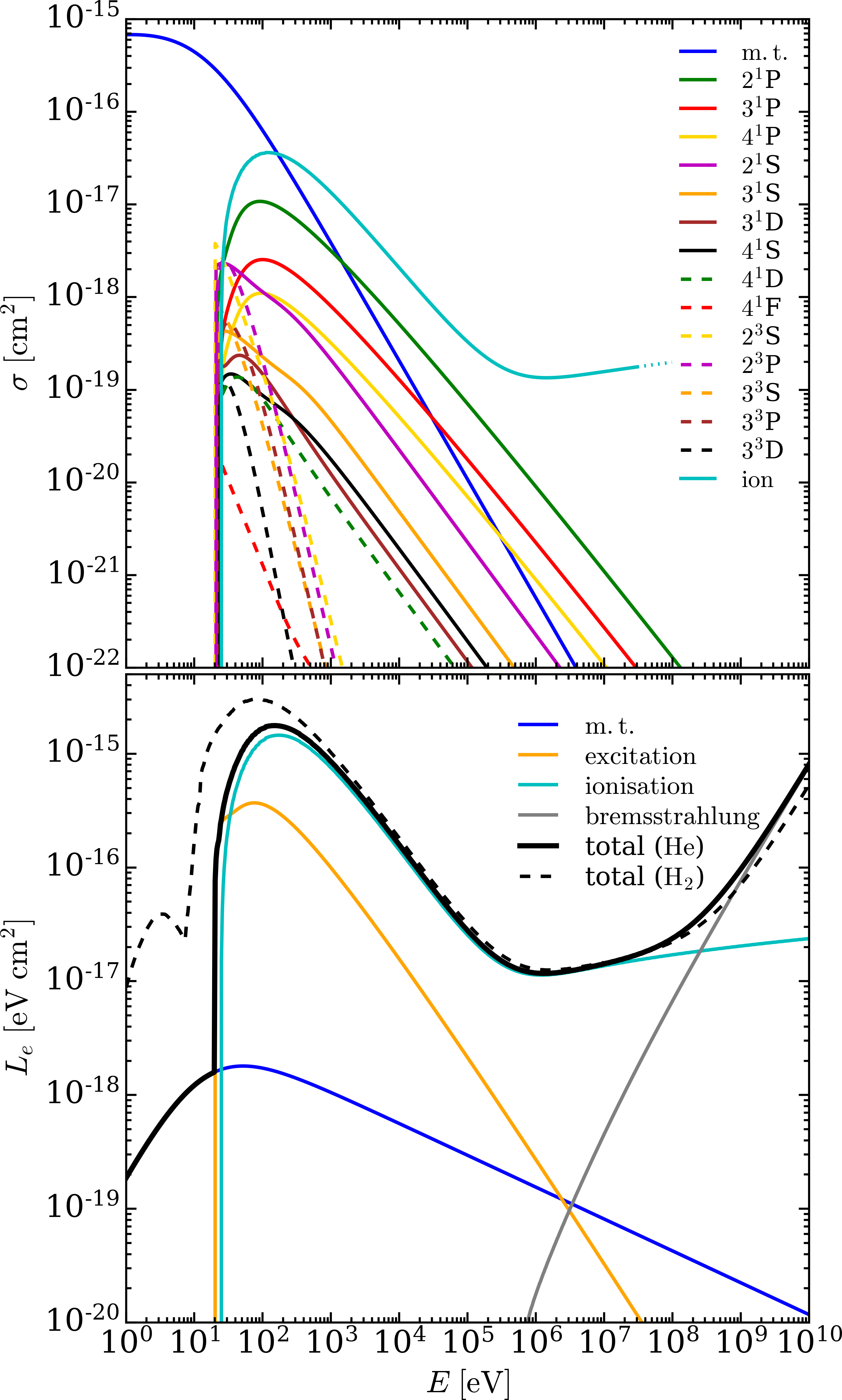}}
\caption{Upper panel:
momentum transfer cross section (``m.t.''; \citealt{PintoGalli2008}), 
excitation cross sections \citep{Ralchenko+2008},
and ionisation cross section (``ion''; \citealt{Kim+2000})
for electrons colliding with He atoms.
Lower panel: energy loss function for electrons colliding with He (solid black line);
momentum transfer loss (``m.t.''; solid blue line),
total excitation loss (solid orange line),
ionisation loss (solid cyan line),
and bremsstrahlung loss (solid grey line, from \citealt{BlumenthalGould1970}).
For comparison, the loss function for electrons colliding with H$_{2}$ (dashed black line) is shown.}
\label{fig:eHe}
\end{center}
\end{figure}

The upper panel of Fig.~\ref{fig:eHe} summarises the excitation and ionisation 
cross sections that we use to derive the energy loss function for electrons colliding 
with He atoms. The equation for calculating the loss function is identical to Eq.~(\ref{eq:lossfunction}), 
except for the pre-factor of the momentum transfer term, where $m_{\rm H_2}$ is replaced by $m_{\rm He}$.
In the lower panel of the same figure we compare the H$_{2}$ and He energy loss functions. 
We note that, by considering a medium with $\sim20\%$ of He, the He loss function has to be 
divided by a factor of $\sim5$.

\section{Differential contribution to the cosmic-ray ionisation rate}
\label{app:EdzetadE}

\begin{figure}[!h]
\begin{center}
\resizebox{\hsize}{!}{\includegraphics[]{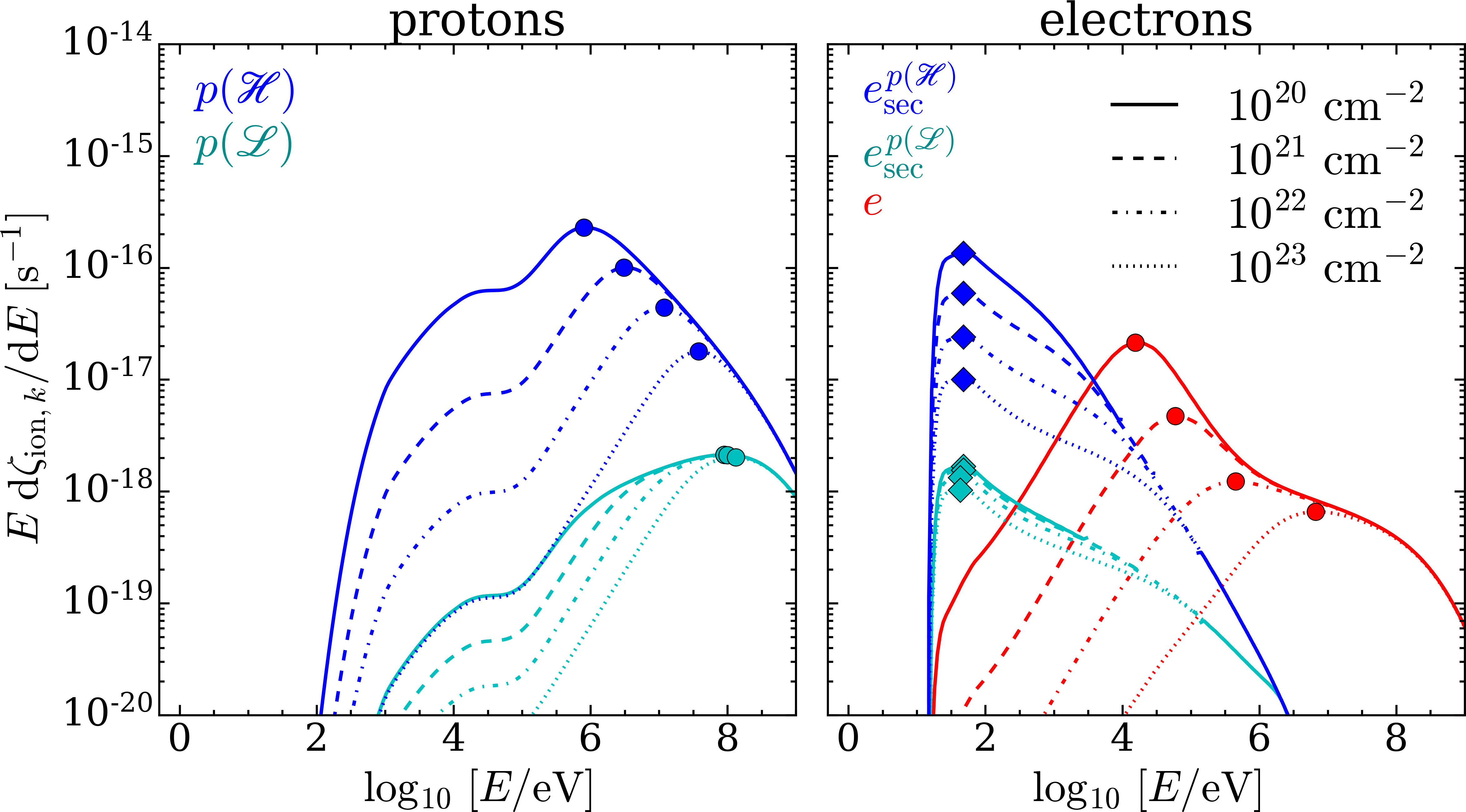}}
\caption{
Differential contribution to the ionisation rate, $E\ud\zeta_{{\rm ion},k}/{\rm d}E$,
per logarithmic energy interval as a function of the energy
at the column densities  $N_{\rm H_{2}}=10^{20}$, $10^{21}$, $10^{22}$, and $10^{23}$~cm$^{-2}$. 
Left plot: CR protons (model $\mathscr{L}$ and $\mathscr{H}$; cyan and blue lines, respectively);
right plot: CR primary electrons (red lines) and
secondary electrons (model $\mathscr{L}$ and $\mathscr{H}$; cyan and blue lines, respectively).
Solid circles (diamonds) 
denote the energies of primary CRs (secondary electrons)
that contribute most to the CR ionisation rate.
}
\label{fig:EdzetadE}
\end{center}
\end{figure}

In order to understand 
why the spectra of secondaries have a different attenuation with column density 
depending on the primary spectrum, it is useful to introduce the differential contribution 
to the ionisation rate per logarithmic energy interval,
$E\ud\zeta_{{\rm ion},k}/\ud E$, where $k$ is the CR species.
This quantity gives an indication of the energy from which the bulk 
of the ionisation is generated \citep[see also][]{Padovani+2009}.
Solid circles in Fig.~\ref{fig:EdzetadE}, 
which are also displayed at the same energies in the lower left panel of 
Fig.~\ref{fig:spectra},
show the primary CR energies that contribute most to 
the CR ionisation rate. 
Accordingly, solid diamonds in the right panel of Fig.~\ref{fig:EdzetadE} refer to secondary electron energies (see also the lower right panel in Fig.~\ref{fig:spectra}).
These energies correspond to the maxima of $E\ud\zeta_{{\rm ion},k}/\ud E$. 
Looking at the left panel of Fig.~\ref{fig:EdzetadE}, we see that
for model $\mathscr{L}$ the peak of $E\ud\zeta_{{\rm ion},p}/\ud E$
is essentially independent of
column density, and its maximum is at 
$E\simeq100$~MeV. 
Conversely, for model $\mathscr{H}$, the peak of $E\ud\zeta_{{\rm ion},p}/\ud E$ 
decreases 
by more than one order of magnitude 
for H$_{2}$ column densities from $10^{20}$~cm$^{-2}$ to $10^{23}$~cm$^{-2}$, and 
its maximum shifts from $E\simeq1$~MeV to $\simeq40$~MeV.
This is because model $\mathscr{H}$ has a non-negligible component of protons at low energies, 
which contribute to the CR ionisation rate. 
However, for increasing column densities, this low-energy tail is quickly attenuated 
\citep{Padovani+2018a},
and thus the peak of $E\ud\zeta_{{\rm ion},p}/\ud E$ moves towards higher energies. 
In contrast, for model $\mathscr{L}$, the largest contribution comes from the 100~MeV protons. 
Such protons are only attenuated at $N_{\rm H_{2}}\gtrsim10^{24}$~cm$^{-2}$, namely 
at column densities outside the range of our interest.
As a result, the secondary electron spectrum from the proton model $\mathscr{L}$ is nearly independent of 
column density, 
while the spectrum from model $\mathscr{H}$ is attenuated at higher column densities.
This is the reason why $\zeta_{{\rm exc},u}$ and
$\zeta_{\rm ion}$ 
for model $\mathscr{L}$ 
show a weak dependence on $N_{\rm H_{2}}$, 
whereas for model $\mathscr{H}$ the dependence is strong 
(see Fig.~\ref{fig:zexc_zion_vs_N}). 
The same reasoning applies to the spectrum of primary electrons for which 
$E\ud\zeta_{{\rm ion},e}/\ud E$ decreases by more than one order of magnitude
for H$_{2}$ column densities from $10^{20}$~cm$^{-2}$ to $10^{23}$~cm$^{-2}$, and 
its maximum shifts from $E\simeq10$~keV to $\simeq10$~MeV.

\section{Cosmic-ray ionisation rate estimates: update from observations}
\label{app:CRionobstheory}
\begin{figure}[!h]
\begin{center}
\resizebox{1\hsize}{!}{\includegraphics[]{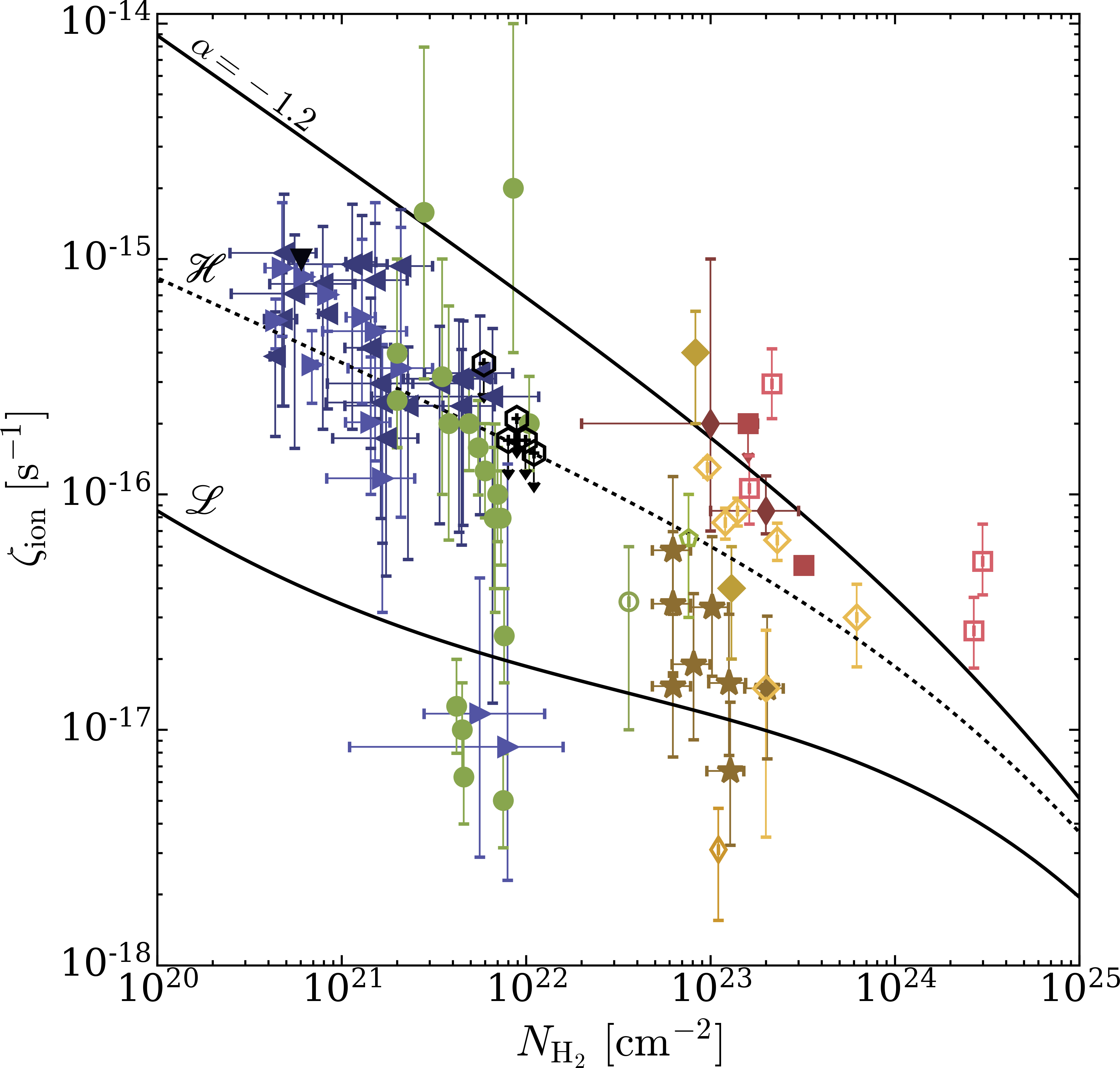}}
\caption{Total CR ionisation rate as a function of the H$_{2}$ column density. 
Theoretical models $\mathscr{L}$ (solid black line), $\mathscr{H}$ (dotted black line),
and with low-energy spectral slope $\alpha=-1.2$ (solid black line).
Expected values from models also include the ionisation due to primary CR electrons and secondary
electrons.
Observational estimates in diffuse clouds:
down-pointing triangle \citep{Shaw+2008},
left-pointing triangles \citep{IndrioloMcCall2012},
right-pointing triangles \citep{NeufeldWolfire2017};
in low-mass dense cores:
solid circles \citep{Caselli+1998},
empty hexagons \citep{Bialy+2022},
empty circle \citep{MaretBergin2007},
empty pentagon \citep{Fuente+2016};
in high-mass star-forming regions:
stars \citep{Sabatini+2020},
solid diamonds \citep{deBoisanger+1996},
empty diamonds \citep{VanderTak+2000},
empty thin diamonds \citep{Hezareh+2008},
solid thin diamonds \citep{MoralesOrtiz+2014};
in circumstellar discs:
solid squares \citep{Ceccarelli+2004};
in massive hot cores:
empty squares \citep{BargerGarrod2020}.
}
\label{fig:zvsN}
\end{center}
\end{figure}

In Fig.~\ref{fig:zvsN} 
we present the estimates of the CR ionisation rate obtained from observations 
in diffuse clouds, low- and high-mass star-forming regions, circumstellar discs, and massive hot cores.
In the same plot we show the trend of $\zeta_{\rm ion}$ predicted by CR propagation models 
\citep[e.g.][]{Padovani+2009,Padovani+2018a}: 
the model $\mathscr{L}$, with low-energy spectral slope $\alpha=0.1$, which is based on the data
of the two Voyager spacecrafts \citep{Cummings+2016,Stone+2019}; 
the model $\mathscr{H}$, with $\alpha=-0.8$, which
reproduces the average value of $\zeta_{\rm ion}$ in diffuse regions; 
the model with $\alpha=-1.2$, which
can be considered as an upper limit to the CR ionisation rate estimates in diffuse regions.
Models also include the contribution of primary CR electrons and secondary electrons.

The spread of $\zeta_{\rm ion}$ in dense cores \citep{Caselli+1998} is supposed to be 
related to uncertainties in the chemical network, in the depletion process of elements such as carbon
and oxygen, as well as because of the presence of tangled magnetic fields 
\citep{PadovaniGalli2011,Padovani+2013,Silsbee+2018}. 
We note that the models presented here only account for the propagation of 
interstellar CRs, 
but in more evolved sources, such as in high-mass star-forming regions and hot cores, 
there could be a substantial contribution from locally accelerated charged particles 
\citep{Padovani+2015,Padovani+2016,GachesOffner2018,Padovani+2021a}.

\end{document}